\documentstyle[amsmath,epsf,a4,11pt]{article}

\date{}

\begin{document}

\title{{\bf Cosmological Perturbations}\\{\small {\em 1st Aegean Summer
School on Cosmology, 21-29 September, 2001, Karlovassi (Samos),
Greece}}}

\author{Christos G. Tsagas\thanks{e-mail address:
ctsagas@maths.uct.ac.za}\\{\small Department of Mathematics and
Applied Mathematics, University of Cape Town}\\ {\small Rondebosch
7701, South Africa}}

\maketitle

\begin{abstract}
The aim of these lecture notes is to familiarize graduate students
and beginning postgraduates with the basic ideas of linear
cosmological perturbation theory and of structure formation
scenarios. We present both the Newtonian and the general
relativistic approaches, derive the key equations and then apply
them to a number of characteristic cases. The gauge problem in
cosmology and ways to circumvent it are also discussed. We outline
the basic framework of the baryonic and the non-baryonic structure
formation scenarios and point out their strengths and
shortcomings. Fundamental concepts, such as the Jeans length, Silk
damping and collisionless dissipation, are highlighted and the
underlying mathematics are presented in a simple and
straightforward manner.
\end{abstract}

\section{Introduction}
Looking up into the night sky we see structure everywhere. Star
clusters, galaxies, galaxy clusters, superclusters and voids are
evidence that on small and moderate scales, that is up to 10~Mpc,
our universe is very lumpy. As we move to larger and larger
scales, however, the universe seems to smooth out. This is
evidenced by the isotropy of the x-ray background, the number
counts of radio sources and, of course, by the high isotropy of
the Cosmic Microwave Background (CMB) radiation. The latter also
provides a fossil record of our observable universe when it was
roughly $10^5$ years old and about $10^3$ times smaller than
today. So, the universe was very smooth at early times and it is
very lumpy now. How did this happen? Although the details are
still elusive, cosmologists believe that the reason is
``gravitational instability''. Small fluctuations in the density
of the primeval cosmic fluid that grew gravitationally into the
galaxies, the clusters and the voids we observe today. The idea of
gravitational instability is not new. It was first introduced in
the early 1900s by Jeans, who showed that a homogeneous and
isotropic fluid is unstable to small perturbations in its
density~\cite{J}. What Jeans demonstrated was that density
inhomogeneities grow in time when the pressure support is weak
compared to the gravitational pull. In retrospect, this is not
surprising given that gravity is always attractive. As long as
pressure is negligible, an overdense region will keep accreting
material from its surroundings, becoming increasingly unstable
until it eventually collapses into a gravitationally bound object.
Jeans, however, applied his analysis to a static Newtonian fluid
in an attempt to understand the formation of planets and stars. In
modern cosmology we need to account for the expansion of the
universe as well as for general relativistic effects.

Despite the lack, as yet, of a detailed scenario, the rather
simple idea that the observed structure in our universe has
resulted from the gravitational amplification of weak primordial
fluctuations seems to work remarkably well. These small
perturbations grew slowly over time until they were strong enough
to separate from the background expansion, turn around, and
collapse into gravitationally bound systems like galaxies and
galaxy clusters. As long as these inhomogeneities are small they
can be studied by the linear perturbation theory. A great
advantage of the linear regime is that the different perturbative
modes evolve independently and therefore can be treated
separately. In this respect, it is natural to divide the analysis
of cosmological perturbations into two regimes. The early phase,
when the perturbation is still outside the horizon, and the late
time regime when the mode is inside the Hubble radius. In the
first case microphysical processes, such as pressure effects for
example, are negligible and the evolution of the perturbation is
basically kinematic. After the mode has entered the horizon,
however, one can no longer disregard microphysics and damping
effects.

In these lectures we will a priori assume the existence of small
inhomogeneities at some initial time in the early universe. A
cosmological model is not complete, however, unless it can also
produce these seed fluctuations through some viable physical
process. Inflation (Guth (1981); Linde (1982)~\cite{G}) appears to
be our best option as it naturally produces a spectrum of
scale-invariant gaussian perturbations. Topological defects
(Kibble (1976)~\cite{K}), such as cosmic strings, global monopoles
and textures, offer a radically different paradigm to inflation
for structure formation purposes. They have fallen out of favor,
however, as their observational situation looks unpromising.

Understanding the details of structure formation requires, among
other things, knowledge of the initial data. Structure formation,
or galaxy formation as it is sometimes referred to, began
effectively with the end of the radiation era at matter-radiation
equality ($t_{\rm eq}\simeq4.4\times10^{10}(\Omega h^2)^{-2}\,{\rm
sec}$, where $\Omega$ is the density parameter of the universe).
Thus, the start of the matter era also signals the beginning of
structure formation. If we were ever to find out the details of
how the structure in our universe formed we need to know the
initial data at that epoch. The necessary information includes:
(i) the total amount of the non-relativistic matter; (ii) the
composition of the universe and the contribution of its various
components to the total density, namely $\Omega_{\rm b}$ from
baryons, $\Omega_{\gamma}$ from relativistic particles,
$\Omega_{\rm WIMP}$ from relic WIMPs,\footnote{Weakly Interacting
Massive Particles (WIMPs) are stable non-baryonic species left
over from the earliest moments of the universe.}
$\Omega_{\Lambda}$ from a potential cosmological constant etc;
(iii) the spectrum and the type (i.e.~adiabatic or isothermal) of
the primeval density perturbations. Given these one can, in
principle, construct a detailed scenario of structure formation,
which then will be tested against observations. The importance of
specifying the initial conditions is paramount, since inverting
present observations to infer the initial data is unfeasible after
all the astrophysical filtering that has taken place. Speculation
on the history of the early universe, backed by recent
observations provide some ``hints'' as to the appropriate initial
data. They point towards $\Omega=1$ from inflation; $\Omega_{\rm
WIMP}\simeq0.3$ (including a small baryonic contribution) and
$\Omega_{\Lambda}\simeq0.7$ from nucleosynthesis, inflation and
the supernovae redshift measurements; and adiabatic fluctuations
with a Harrison-Zeldovich spectrum~\cite{H} from inflation.

The layout of these notes is as follows. In Sec.~2 we present the
Newtonian treatment of linear cosmological perturbations, discuss
issues such as pressure support and the ``Jeans length'' and
provide the key results. The general relativistic analysis is
outlined in Sec.~3 and the basic linear equations are derived. We
also give a brief discussion of the ``gauge problem'' and provide
characteristic solutions of the relativistic approach. In Sec.~4
we discuss entirely baryonic structure formation scenarios,
emphasizing the collisional damping of adiabatic density
perturbations. Non-baryonic ``hot'' and ``cold'' dark matter
models are presented in Sec.~5, together with their advantages and
shortcomings. The aim of these lectures notes is to provide the
basic background to graduate students in physics and astronomy as
well as to beginning postgraduates. We would like to familiarize
the newcomer with fundamental concepts such as gravitational
instability, the Jeans length, collisional and collisionless
damping. The necessary mathematics are also provided in simple and
straightforward manner. Overall we want to give a brief but
comprehensive picture of the linear regime, so that the interested
student will feel more confident when looking at more
sophisticated treatments. For further details we refer the reader
to some of excellent monographs that now exist in the literature
(see~\cite{KT} for a list of them). The lectures do not require a
particularly specialized background, although some knowledge of
cosmology and general relativity will be helpful.

\section{Linear Newtonian perturbations}
The Newtonian theory, as a limiting approximation of general
relativity, is only applicable to scales well within the Hubble
radius where the effects of spacetime curvature are
negligible.\footnote{Throughout these lectures we will de dealing
with cosmological models where the Hubble radius and the particle
horizon, that is the maximum proper distance travelled by a
typical photon, are effectively identical. We will therefore use
the two concepts interchangeably. The reader is referred to
Padmanabhan (1993)~\cite{KT} for an illuminating discussion on the
differences between these two important cosmological scales.} Even
in this context, however, one can only analyze density
perturbations in the non-relativistic component. Perturbations in
the relativistic matter, at all scales, require the full theory.
After equipartition the universe is dominated by non-relativistic
pressureless matter, which is commonly referred to as ``dust''.
Thus, it becomes clear that the Newtonian analysis applies only to
late times in the lifetime of our universe. For earlier times and
larger scales one needs to employ general relativity. Curiously
enough, it was not until 1957, long after Lifshitz's fully
relativistic treatment, that Bonnor employed the Newtonian theory
to study perturbations in a dust dominated FRW cosmology~\cite{B}.
In some ways the relativistic approach is simpler than the
Newtonian, which requires considerable mathematical subtlety. For
an extensive discussion of the Newtonian approach, we refer the
reader to Peebles (1980)~\cite{W}. Here, we will apply the idea of
the Jeans instability to the simple case of an expanding
self-gravitating non-relativistic fluid.

\subsection{The general fluid equations}
To begin with let us consider some fundamental ideas that apply to
both Newtonian and relativistic settings. We model the universe as
a fluid, so that all the relevant quantities are described by
smoothly varying functions of position. Cosmic strings, domain
walls and other topological defects have no place in this picture.
After the first bound structures, such as galaxies, form they are
treated like particles along with the genuine particles that
remain unbound. Note that such a fluid description of the universe
applies only after smoothing over large comoving scales. An
additional important concept is that of the ``comoving observer''.
Loosely speaking, a comoving observer follows the expansion of the
universe including the effects of any inhomogeneities that may be
present. Adopting a Newtonian reference frame, specified by
Cartesian space coordinates $r_{\alpha}$ and a universal time $t$,
we consider a fluid with density $\rho$ and pressure $p$, moving
with velocity $v_{\alpha}$ in a gravitational potential
$\Phi$.\footnote{Greek indices indicate Newtonian spaces and take
the values 1,2,3, while Latin characters run between 0 and 3 and
denote spacetime quantities.} Its evolution is governed by the
standard Eulerian equations for a self gravitating medium
\begin{eqnarray}
\frac{\partial\rho}{\partial t}+ \partial_{\alpha}(\rho
v^{\alpha})&=&0\,,  \label{Neqp1}\\ \frac{\partial
v_{\alpha}}{\partial t}+ v^{\beta}\partial_{\beta}v_{\alpha}+
\frac{1}{\rho}\partial_{\alpha}p+
\partial_{\alpha}\Phi&=&0\,, \label{Neqp2}\\
\partial^2\Phi-4\pi G\rho&=&0\,.  \label{Neqp3}
\end{eqnarray}
where $\partial^2=\partial^{\alpha}\partial_{\alpha}$ is the
associated Laplacian operator. Expressions
(\ref{Neqp1})-(\ref{Neqp3}) are respectively known as the
continuity, the Euler and the Poisson equations. They describe
mass conservation, momentum conservation and the Newtonian
gravitational potential. The simplest solution to the above system
is not applicable to cosmology, as it corresponds to a static
matter distribution with $\rho\,,p={\rm constant}$.\footnote{In
such a configuration the gravitational force vanishes
(i.e.~$\partial_{\alpha}\Phi=0$ from Eq.~(\ref{Neqp2})). This
clearly contradicts the Poisson equation, as it seems to suggest
that the matter density vanishes as well. Following Jeans, we
assume that Eq.~(\ref{Neqp3}) describes relations between
perturbed quantities only (Jeans' swindle).} For an expanding
fluid it is convenient to adopt a ``comoving" coordinate set
($x_{\alpha}$) instead of the ``physical" (or proper) coordinates
($r_{\alpha}$) employed in (\ref{Neqp1})-(\ref{Neqp3}). The two
frames are related via
\begin{equation}
r_{\alpha}=ax_{\alpha}\,,  \label{pcc}
\end{equation}
where $a=a(t)$ is the scale factor of the universe. The above
immediately implies the relation
\begin{equation}
v_{\alpha}=Hr_{\alpha}+u_{\alpha}\,,  \label{pcv}
\end{equation}
between the physical velocity $v_{\alpha}={\rm d}r_{\alpha}/{\rm
d}t$ and the ``peculiar" velocity $u_{\alpha}={\rm
d}x_{\alpha}/{\rm d}t$, where $H=\dot{a}/a$ represents the Hubble
parameter. In a comoving frame Eqs.~(\ref{Neqp1})-(\ref{Neqp3})
read
\begin{eqnarray}
\frac{\partial\rho}{\partial t}+ 3H\rho+
\frac{1}{a}\partial_{\alpha}(\rho u^{\alpha})&=&0\,,  \label{Neqc1}\\
\frac{{\rm d}^2a}{{\rm d}t^2}x_{\alpha}+ \frac{\partial
u_{\alpha}}{\partial t}+ Hu_{\alpha}+
\frac{1}{a}u^{\beta}\partial_{\beta}u_{\alpha}+
\frac{1}{a\rho}\partial_{\alpha}p+
\frac{1}{a}\partial_{\alpha}\Phi&=&0\,,  \label{Neqc2}\\
\partial^2\Phi-4\pi Ga^2\rho&=&0\,.  \label{Neqc3}
\end{eqnarray}
One arrives at the above from (\ref{Neqp1})-(\ref{Neqp3}) on using
the transformation laws $(\partial/\partial t)_{\rm
phys}=(\partial/\partial t)_{\rm com}-
Hx^{\alpha}\partial_{\alpha}$ and $(\partial_{\alpha})_{\rm
phys}=(1/a)(\partial_{\alpha})_{\rm com}$ between physical and
comoving derivatives.

\subsection{The unperturbed background}
The simplest non-static solution to the system
(\ref{Neqc1})-(\ref{Neqc3}) describes a smoothly expanding
(i.e.~$u_{\alpha}=0$), homogeneous and isotropic fluid
(i.e.~$\rho_0=\rho_0(t)\,,~p_0=p_0(t)$). In particular, the
unperturbed background universe is characterized by the system
\begin{eqnarray}
\frac{{\rm d}\rho_0}{dt}+ 3H\rho_0&=&0\,,  \label{Neqb1}\\
\frac{{\rm d}^2a}{dt^2}x_{\alpha}+
\frac{1}{a}\partial_{\alpha}\Phi_0&=&0\,,
\label{Neqb2}\\
\partial^2\Phi_0- 4\pi Ga^2\rho_0&=&0\,,  \label{Neqb3}
\end{eqnarray}
with solutions
\begin{equation}
\rho_0\propto
a^{-3}\,,\hspace{5mm}v^{\alpha}_0=Hr_{\alpha}\hspace{5mm}{\rm
and}\hspace{5mm}\Phi_0={\textstyle{2\over3}}\pi Ga^2\rho_0x^2\,.,
\label{Nbs}
\end{equation}
where the expression for the gravitational potential follows from
the isotropy assumption, namely from the fact that
$\partial_{\alpha}=\partial/\partial x_{\alpha}={\rm d}/{\rm d}x$,
and equation  ${\rm d}^2a/{\rm d}t^2=-4\pi Ga\rho/3$. Note that we
no longer need to invoke Jeans' swindle, although the approach is
still not entirely problem free, as $\Phi_0$ has a spatial
dependence despite the assumption of spatial homogeneity. Also, on
large scales $r_{\alpha}>1/H\sim\lambda_{H}$, which substituted
into Eq.~(\ref{Nbs}b) gives an expansion velocity greater than
that of light. We remind the reader, however, that the Newtonian
treatment applies to sub-horizon scales only.

\subsection{The linear regime}
Consider perturbations about the aforementioned background
solution
\begin{equation}
\rho=\rho_0+\delta\rho\,, \hspace{5mm} p=p_0+\delta p\,,
\hspace{5mm} v^{\alpha}=v_0^{\alpha}+\delta v^{\alpha}\,,
\hspace{5mm} \Phi=\Phi_0+\delta\Phi\,.  \label{deltas}
\end{equation}
where $\delta\rho$, $\delta p$, $\delta v^{\alpha}$ and
$\delta\Phi$ are the perturbed first order variables with spatial
as well as temporal dependence
(i.e.~$\delta\rho=\delta\rho(t,x_{\alpha}))$. In the linear regime
the perturbed quantities are much smaller than their zero order
counterparts (i.e.~$\delta\rho\ll\rho_0$) During this period
higher order terms, for example the product $\delta\rho\delta
v_{\alpha}$, are negligible. This means that different
perturbative modes evolve independently and therefore can be
treated separately. Note that $\delta v_{\alpha}\equiv u_{\alpha}$
is simply the peculiar velocity describing deviations from the
smooth Hubble expansion. Also, the fluid pressure is related to
the density via the equation of state of the medium. For
simplicity, we will only consider ``barotropic'' fluids with
$p=p(\rho)$.

Substituting (\ref{deltas}) into Eq.~(\ref{Neqc1}) and keeping up
to first order terms only we obtain
\begin{eqnarray}
\frac{\delta\rho}{\partial t}+ 3H\delta\rho+
\frac{\rho_0}{a}\partial_{\alpha}\delta v^{\alpha}=0\,,
\label{ldeltarho}
\end{eqnarray}
on using the zero-order expression ${\rm d}\rho_0/{\rm
d}t+3H\rho_0=0$ (see Eq.~(\ref{Neqb1})). The above describes the
linear evolution of density fluctuations. In perturbation
analysis, however, it is advantageous to employ dimensionless
variables for first order quantities. Here, we will be using the
dimensionless ``density contrast" $\delta\equiv\delta\rho/\rho_0$.
Throughout the linear regime $\delta\rho\ll\rho_0$ meaning that
$\delta\ll1$. Introducing $\delta$ we may recast
Eq.~(\ref{ldeltarho}) into
\begin{equation}
\frac{\partial\delta}{\partial t}+
\frac{1}{a}\partial_{\alpha}\delta v^{\alpha}=0\,,  \label{ldelta}
\end{equation}
given that $\partial\delta/\partial
t=(1/\rho_0)\partial\delta\rho/\partial t+(3H/\rho_0)\delta\rho$
from Eq.~(\ref{Neqb1}). Also, substituting (\ref{deltas}) into
Eq.~(\ref{Neqc2}) and linearizing we obtain the propagation
equation for the velocity perturbation\footnote{Velocity
perturbations decompose into rotational modes $\delta
v_{\perp}^{\alpha}$, where $\partial_{\alpha}\delta
v_{\perp}^{\alpha}=0$, and irrotational modes $\delta
v_{\parallel}^{\alpha}$ with ${\rm curl}\delta
v_{\parallel}^{\alpha}=0$. However, it is only the divergence of
$\delta v^{\alpha}$ (i.e.~$\partial_{\alpha}\delta v^{\alpha}$)
which contributes to the evolution of the density contrast (see
Eq.~(\ref{ldeltarho})). Thus, rotational modes do not couple to
the longitudinal density perturbations we will be addressing in
these lectures.}
\begin{eqnarray}
\frac{\partial\delta v_{\alpha}}{\partial t}+ H\delta v_{\alpha}+
\frac{v_{\rm s}^2}{a\rho_0}\partial_{\alpha}\delta+
\frac{1}{a}\partial_{\alpha}\delta\Phi=0\,.  \label{ldeltavi}
\end{eqnarray}
In deriving the above we have used the zero-order part of
Eq.~(\ref{Neqc2}) and the binomial law
$(1+\delta)^{-1}\simeq1-\delta$ (recall that $\delta\ll1$). Also,
for a barotropic fluid $p=p(\rho)$ and $\partial_{\alpha}\delta
p=v_{\rm s}^2\partial_{\alpha}\delta\rho$, where $v_{\rm
s}^2\equiv dp/d\rho$ is the adiabatic sound speed. Finally,
inserting (\ref{deltas}) into the Poisson equation and linearizing
we arrive at the relation
\begin{equation}
\partial^2\delta\Phi- 4\pi Ga^2\rho_0\delta=0\,,
\label{ldeltaPhi}
\end{equation}
for the evolution of the perturbed gravitational potential.
Results (\ref{ldelta})-(\ref{ldeltaPhi}) determine the behavior of
the density perturbation completely. When combined they lead to a
second order differential equation that describes the linear
evolution of the density contrast. In particular, taking the time
derivative of (\ref{ldelta}) and employing Eqs.~(\ref{ldeltavi}),
(\ref{ldeltaPhi}) we find that
\begin{eqnarray}
\frac{\partial^2\delta}{\partial t^2}- \frac{v_{\rm
s}^2}{a^2}\partial^2\delta=- 2H\frac{\partial\delta}{\partial t}+
4\pi G\rho_0\delta\,, \label{Ndel}
\end{eqnarray}
to linear order. We have arrived at the above by assuming
Newtonian gravity, which requires matter domination and zero
cosmological constant. As long as we remain well within the
horizon, however, Eq.~(\ref{Ndel}) also describes ordinary matter
perturbations in the presence of a smooth radiative background or
a cosmological constant. This is so because general relativity
always reduces to Newtonian gravity near a free-falling observer
moving at non-relativistic speeds. The only difference (see
Peebles (1980)~\cite{W}) is that the Poisson equation has been
replaced by Eq.~(\ref{EFE1}). This changes the smooth
gravitational potential but not its perturbation, provided the
latter is due to matter alone. As a result, Eq.~(\ref{Ndel}) is
still valid if $\rho$ is the simply density of the
non-relativistic matter and $\delta$ the associated density
contrast.

\subsection{The Jeans Length}
Equation (\ref{Ndel}) is a wave-like equation with two extra terms
in the right-hand side; one due to the expansion of the universe
and the other due to gravity. It is therefore, natural to seek
plane wave solutions of the form
\begin{equation}
\delta=\sum_{\rm k}\tilde{\delta}_{({\rm k})}{\rm e}^{i{\rm
k}_{\alpha}x^{\alpha}}\,, \label{Fourier}
\end{equation}
where $\tilde{\delta}_{({\rm k})}=\tilde{\delta}_{({\rm k})}(t)$,
${\rm k}_{\alpha}$ is the comoving wavevector and ${\rm
k}=\sqrt{{\rm k}_{\alpha}{\rm k}^{\alpha}}$ is the associated
comoving wavenumber.\footnote{The comoving wavelength of the
perturbative mode is given by $\ell=2\pi/{\rm k}$, while
$\lambda=a\ell$ is the physical (proper) wavelength. Also, the
physical wavenumber is ${\rm n}=2\pi/\lambda$ with ${\rm n}={\rm
k}/a$. Finally, we note that ${\rm k}_{\alpha}x^{\alpha}={\rm
n}_{\alpha}r^{\alpha}$.} Fourier decomposing Eq.~(\ref{Ndel}) and
using the relations $\partial_{\alpha}\delta=i{\rm
k}_{\alpha}\delta$, $\partial^2\delta=-{\rm k}^2\delta$ we obtain
\begin{equation}
\frac{{\rm d}^2\tilde{\delta}_{({\rm k})}}{{\rm d}t^2}=
-2H\frac{{\rm d}\tilde{\delta}_{({\rm k})}}{{\rm d}t}+ \left(4\pi
G\rho_0- \frac{v_{\rm s}^2{\rm
k}^2}{a^2}\right)\tilde{\delta}_{({\rm k})}\,, \label{Ndelk}
\end{equation}
which determines the evolution of the ${\rm k}$-th perturbative
mode. The first term in the right-hand side of the above is due to
the expansion and always suppresses the growth of
$\tilde{\delta}_{({\rm k})}$. The second reflects the conflict
between pressure support and gravity. When $4\pi G\rho_0\gg v_{\rm
s}^2k^2/a^2$ gravity dominates. On the other hand, pressure
support wins if $v_{\rm s}^2k^2/a^2\gg4\pi G\rho_0$. The threshold
$4\pi G\rho_0=v_{\rm s}^2k^2/a^2$ defines the length scale
\begin{equation}
\lambda_{\rm J}=v_{\rm s}\sqrt{\frac{\pi}{G\rho_0}}\,. \label{Jl}
\end{equation}
The physical scale $\lambda_{\rm J}$, known as the ``Jeans
length'', constitutes a characteristic feature of the
perturbation. It separates the gravitationally stable modes from
the unstable ones. Fluctuations on scales well beyond
$\lambda_{\rm J}$ grow via gravitational instability, while modes
with $\lambda\ll\lambda_{\rm J}$ are stabilized by
pressure.\footnote{One also arrives at the Jeans length following
a simple qualitative argument. The timescale for gravitational
collapse is $t_{\rm grav}\sim(1/G\rho)^{1/2}$, whereas the
timescale for the pressure forces to respond is $t_{\rm
press}\sim\lambda/v_{\rm s}$. When $t_{\rm grav}\ll t_{\rm
press}$, that is when $\lambda\gg v_{\rm
s}(1/G\rho)^{1/2}\sim\lambda_{\rm J}$, pressure gradients do not
have the time to respond and restore hydrostatic equilibrium. In
the opposite case, namely for $t_{\rm press}\ll t_{\rm grav}$,
pressure totally overwhelms gravity.} The Jeans length corresponds
to the ``Jeans mass", defined as the mass contained within a
sphere of radius $\lambda_{\rm J}/2$
\begin{equation}
M_{\rm J}={\textstyle{4\over3}}\pi\rho\left(\frac{\lambda_{\rm
J}}{2}\right)^3\,, \label{Jm}
\end{equation}
where $\rho$ is the density of the perturbed component.

\subsection{Multi-component fluids}
Thus far we have only considered a single-component fluid. When
dealing with a multi-component medium (e.g.~baryons, photons,
neutrinos or other exotic particles), perturbations in the
non-relativistic component evolve according to
\begin{equation}
\frac{{\rm d}^2\delta_{i}}{{\rm d}t^2}=-2H\frac{{\rm
d}\delta_{i}}{{\rm d}t}+ \left[4\pi
G\rho_0\sum_{j}\epsilon_{j}\delta_{j}-\frac{\left(v_{\rm
s}^2\right)_{i}{\rm k}^2}{a^2}\delta_{i}\right]\,, \label{mcNdel}
\end{equation}
where the index $i$ refers to the component in question. The sum
is over all species and $\epsilon_{i}=\rho_{i}/\rho_0$ provides a
measure of each component's contribution to the total background
density $\rho_0=\sum_{i}\rho_{i}$. Note that any smoothly
distributed matter does not contribute to the right-hand side of
the above. However, the unperturbed density of this component
contributes to the background expansion. To first approximation,
$H$ is determined by the component that dominates gravitationally,
while $v_{\rm s}$ is the velocity dispersion of the perturbed
species which provide the pressure support.\footnote{The pressure
of a baryonic gas is is the result of particle collisions. For
``dark matter'', collisions are negligible and pressure support
comes from the readjustment of the orbits of the colisionless
species. In both cases it is the velocity dispersion of the
perturbed component which determines the associated Jeans length.}
Equation (\ref{mcNdel}) applies to the ${\rm k}$-th perturbative
mode, although we have omitted the wavenumber (and the tilde) for
simplicity.

\subsection{Solutions}
We will now look for solutions to Eqs.~(\ref{Ndelk}),
(\ref{mcNdel}) in the following four different situations:
(i)~Perturbations in the dominant non-relativistic component
(baryonic or not) for $t>t_{\rm eq}$. (ii)~Fluctuations in the
non-baryonic matter for $t<t_{\rm eq}$. (iii)~Baryonic
perturbations in the presence of a dominant collisionless species.
(iv)~Perturbations in the matter component during a late-time
curvature dominated regime.

\subsubsection{Perturbed Einstein - de Sitter universe}
Consider a dust dominated (i.e.~$p=0=v_{\rm s}^2$) FRW cosmology
with flat spatial sections (i.e.~$\Omega=1$). This model, also
known as the Einstein-de Sitter universe, is though to provide a
good description our universe after recombination. To zero order
$a\propto t^{2/3}$, $H=2/3t$ and $\rho_0=1/6\pi Gt^2$. Perturbing
this background, we look at scales well bellow the Hubble radius
where the Newtonian treatment is applicable. On using definition
(\ref{Jl}) and the relation $\lambda=2\pi a/{\rm k}$,
Eq.~(\ref{Ndelk}) reads
\begin{equation}
t^2\frac{{\rm d}^2\delta}{{\rm d}t^2}+
{\textstyle{4\over3}}t\frac{{\rm d}\delta}{{\rm d}t}-
{\textstyle{2\over3}}\left[1 -\left(\frac{\lambda_{\rm
J}}{\lambda}\right)^2\right]\delta=0\,. \label{EdS}
\end{equation}
Note that form now on we will drop the tilde (${}^{\sim}$) and the
wavenumber (${\rm k}$) for convenience. For modes well within the
horizon but still much larger than the Jeans length (i.e.~when
$\lambda_{\rm J}\ll\lambda\ll\lambda_H$), we find
\begin{equation}
\delta={\cal C}_1t^{2/3}+{\cal C}_2t^{-1}\,, \label{EdS1}
\end{equation}
for the evolution of the density contrast. As expected, there are
two solutions: one growing and one decaying. Any given
perturbation is expressed as a linear combination of the two
modes. At late times, however, only the growing mode is
important.\footnote{Solution (\ref{EdS1}) demonstrates the
difference between the Jeans instability in the static
regime~\cite{J} (e.g.~within a galaxy) and in the expanding
universe. The expansion slows the exponential growth of the static
environment down to a power law.} Therefore, after matter
radiation equality perturbations in the non-relativistic component
grow proportionally to the scale factor (recall that $a\propto
t^{2/3}$ for dust). Note that baryonic perturbations cannot grow
until matter has decoupled from radiation at recombination (we
always assume that $t_{\rm eq}<t_{\rm rec}$). Dark matter
particles, on the other hand, are already collisionless and
fluctuations in their density can grow immediately after
equipartition. After recombination, perturbations in the baryons
also grow proportionally to the scale factor.

On scales well bellow the Jeans length (i.e.~for $\lambda\ll
\lambda_{\rm J}$), Eq.~(\ref{EdS}) admits the solution
\begin{equation}
\delta={\cal C}t^{-1/6}{\rm e}^{\pm i\sqrt{2/3}(\lambda_{\rm
J}/\lambda){\rm ln}t}\,, \label{EdS2}
\end{equation}
which describes a damped oscillation. Thus, small-scale
perturbations in the non-relativistic matter are suppressed by
pressure. Note that $\lambda_{\rm J}/\lambda\propto\nu/H$, where
$\nu=v_{\rm s}/\lambda$ is the oscillation frequency. For
$\lambda_{\rm J}\gg\lambda$ there are many oscillations within an
expansion timescale (adiabatically slow expansion). Also, before
recombination baryons and photons are tightly coupled and $(v_{\rm
s}^2)_{\rm b}\propto T_{\rm b}\simeq T_{\gamma}\propto a^{-1}$,
implying that $\lambda_{\rm J}/\lambda={\rm constant}$. After
decoupling $(v_{\rm s}^2)_{\rm b}\propto T_{\rm b}\propto a^{-2}$.

\subsubsection{Mixture of radiation and dark matter}
Consider the radiation dominated regime when $a\propto t^{1/2}$
and $H=1/2t$. On scales much smaller than the Hubble radius we can
employ the Newtonian theory to study perturbations in the
non-relativistic matter. The Newtonian equations are still
applicable provided that the expansion is determined by the
dominant radiative component. Applying Eq.~(\ref{mcNdel}) to a
mixture of radiation and collisionless particles, with $v_{\rm
s}=0$, we have
\begin{equation}
\frac{{\rm d}^2\delta_{\rm DM}}{{\rm d}t^2}+ 2H\frac{{\rm
d}\delta_{\rm DM}}{{\rm d}t}= 4\pi
G\rho_0\left(\epsilon_{\gamma}\delta_{\gamma}+\epsilon_{\rm
DM}\delta_{\rm DM}\right)\simeq 4\pi
G\rho_{\gamma}\delta_{\gamma}\,, \label{r+dmNdel}
\end{equation}
since $\rho_{\rm DM}\ll\rho_{\gamma}\simeq\rho_0$
(i.e.~$\epsilon_{\rm DM}\ll1$ whereas $\epsilon_{\gamma}\simeq1$).
Given that prior to equipartition $H=1/2t$ and the small-scale
photon distribution is smooth
(i.e.~$\langle\delta_{\gamma}\rangle\simeq0$ see Sec.~3.4.2), the
above reduces to
\begin{equation}
t\frac{{\rm d}^2\delta_{\rm DM}}{{\rm d}t^2}+ \frac{{\rm
d}\delta_{\rm DM}}{{\rm d}t}=0\,, \label{r+dmNdel1}
\end{equation}
and admits the solution
\begin{equation}
\delta_{\rm DM}={\cal C}_1+ {\cal C}_2{\rm ln}t\,. \label{r+dm}
\end{equation}
Thus, in the radiation epoch small scale perturbations in the
collisionless component experience a very slow logarithmic growth,
even when $\lambda>\lambda_{\rm J}$.\footnote{A more careful
treatment, keeping the right-hand side in Eq.~(\ref{r+dmNdel})
gives $\delta_{\rm DM}\propto1+3a/2a_{\rm eq}$ (see Padmanabhan
(1993); Coles \& Lucchin (1995)~\cite{KT}). Hence, $\delta_{\rm
DM}\sim{\rm constant}$ for $a\ll a_{\rm eq}$, namely during the
radiation era. As the dust era progresses $a\gg a_{\rm eq}$ and
$\delta_{\rm DM}\propto a$ in agreement with Sec.~2.6.1.} The
stagnation or freezing-in of matter perturbations prior to
equilibrium is generic to models with a period of rapid expansion
dominated by relativistic particles and is sometimes referred to
as the Meszaros effect~\cite{M}.

\subsubsection{Mixture of dark matter and baryons}
During the period from equilibrium to recombination perturbations
in the dark component grow by a factor of $a_{\rm rec}/a_{\rm
eq}={\rm T}_{\rm eq}/{\rm T}_{\rm rec}\simeq21\Omega h^2$. At the
same time baryonic fluctuations do not experience any growth
because of the tight coupling between photons and baryons. After
decoupling, perturbations in the ordinary matter will also start
growing driven by the gravitational potential of the collisionless
species. To be precise, consider the post-recombination universe,
with $a\propto t^{2/3}$ and $H=2/3t$, dominated by non-baryonic
dark matter. Following Sec.~2.6.1, perturbations in the
collisionless component grow as $\delta_{\rm DM}={\cal C}a$, where
${\cal C}$ is a constant. For baryonic fluctuations on scales
larger that $\lambda_{\rm J}$, Eq.~(\ref{mcNdel}) gives
\begin{equation}
\frac{{\rm d}^2\delta_{\rm b}}{{\rm d}t^2}+ 2H\frac{{\rm
d}\delta_{\rm b}}{{\rm d}t}= 4\pi G\rho_0\left(\epsilon_{\rm
DM}\delta_{\rm DM}+\epsilon_{\rm b}\delta_{\rm b}\right)\simeq
4\pi G\rho_{\rm DM}\delta_{\rm DM}\,, \label{dm+bNdel}
\end{equation}
since $v_{\rm s}\simeq0$ for both species and $\rho_{\rm
b}\ll\rho_{\rm DM}\simeq\rho_0$. Introducing the scale factor as
the independent variable we may recast the above as
\begin{equation}
a^{3/2}\frac{{\rm d}}{{\rm d}a}\left(a^{-1/2}\frac{{\rm
d}\delta_{\rm b}}{{\rm d}a}\right)+ 2\frac{{\rm d}\delta_{\rm
b}}{{\rm d}a}={\textstyle{3\over2}}{\cal C}\,,  \label{dm+bNdel1}
\end{equation}
where we have also used the relation $a\propto t^{2/3}$. The
initial conditions at recombination are $\delta_{\rm b}=0$,
because of the tight coupling between the baryons and the smoothly
distributed photons, and $\delta_{\rm DM}\neq0$ given that the
dark matter particles are already collisionless. The solution
\begin{equation}
\delta_{\rm b}=\delta_{\rm DM}\left(1-\frac{a_{\rm
rec}}{a}\right)\,,  \label{dm+b}
\end{equation}
shows that $\delta_{\rm b}\rightarrow\delta_{\rm DM}$ as $a\gg
a_{\rm rec}$. In other words, baryonic fluctuations quickly catch
up with perturbations in the dark matter component after
decoupling. Alternatively, one might say that the baryons fall
into the ``potential wells'' created by the collisionless species.

\subsubsection{The curvature dominated regime}
So far, and also for the rest of these notes, we have assumed a
high density background universe with $\Omega$ close to unity. If,
instead, $\Omega$ is small the universe could become curvature
dominated at late times. Soon after the transition into the
curvature regime, the rapid expansion of the universe prevents the
inhomogeneities from growing. Let us take a closer look at this
issue. When curvature dominates $\Omega\ll1$, $a\propto t$ and
$H=1/t$ (see Kolb \& Turner (1990)~\cite{KT}). For collisionless
matter with $v_{\rm s}\simeq0$ Eq.~(\ref{Ndelk}) gives
\begin{equation}
t^2\frac{{\rm d}^2\delta}{{\rm d}t^2}+ 2t\frac{{\rm d}\delta}{{\rm
d}t}= {\textstyle{3\over2}}\Omega\delta\,, \label{cdel}
\end{equation}
since $4\pi G\rho_0=3\Omega/2t^2$. Moreover, given that
$\Omega\ll1$ the right-hand side of the above is effectively
second order, and Eq.~(\ref{cdel}) reduces to
\begin{equation}
t^2\frac{{\rm d}^2\delta}{{\rm d}t^2}+ 2t\frac{{\rm d}\delta}{{\rm
d}t}=0\,. \label{cdel1}
\end{equation}
The solution
\begin{equation}
\delta={\cal C}_1+{\cal C}_2t^{-1}\,,  \label{c}
\end{equation}
verifies that perturbations cease growing when curvature
dominates.

\subsection{Summary}
The Newtonian treatment suffices on sub-horizon scales and as long
as we we deal with fluctuations in the non-relativistic component.
During the radiation and the curvature dominated eras,
perturbations do not grow. This suppression occurs because, in
both epochs, the expansion is too rapid for the perturbations to
experience any growth. After equipartition, fluctuations with
wavelengths larger than the Jeans length grow as $\delta\propto
a\propto t^{2/3}$, whereas those bellow $\lambda_{\rm J}$
oscillate like acoustic waves. Note that baryonic perturbations do
not grow until recombination due to the tight coupling between
baryons and photons. Dark matter fluctuations, on the other hand,
start growing immediately after matter-radiation equality. As soon
as the baryons have decoupled perturbations in their density
distribution will be driven by the gravitational potential of the
collisionless species. Thus, shortly after recombination, baryonic
fluctuations grow rapidly and soon equalize with those in the dark
matter. Subsequently, perturbations in both components grow
proportionally to the scale factor.

\section{Linear relativistic perturbations}
Although the Newtonian analysis provides valuable insight into the
behavior of inhomogeneities, it has serious shortcomings. The
proper wavelength of any perturbative mode will be bigger than the
horizon at sufficiently early times. On such scales general
relativistic effects become important and gauge ambiguities need
to be settled. Moreover, one cannot use the Newtonian theory to
study perturbations in the relativistic component. Given that all
the astrophysically relevant scales were outside the horizon early
on (i.e.~for $z>10^3$), it becomes clear that a general
relativistic treatment of cosmological density perturbations is
imperative. General relativity was first applied to cosmological
perturbations in a seminal paper by Lifshitz in 1946~\cite{L}. The
Lifshitz approach relies upon selecting a ``gauge'', finding the
solutions in that gauge and then identifying the ``gauge modes''.
In the relativistic treatment one perturbs both the spacetime
metric and the energy-momentum tensor of the matter sources. In
other words, one assumes that $g_{ab}=g_{ab}^{_0}+\delta g_{ab}$
and $T_{ab}=T_{ab}^{_0}+\delta T_{ab}$, where the zero index
denotes the background variables. For ``small'' $\delta g_{ab}$
and $\delta T_{ab}$, one can perturb and then linearize the
Einstein Field Equations (EFE). Here, we will only outline the key
steps of the analysis and state the main results, while we refer
the reader to~\cite{W,KT} for the details. Note that the complete
set of the relativistic equations reveals three types of
perturbations, namely tensor, vector and scalar modes. Tensor
perturbations correspond to the traceless, transverse part of
$\delta g_{ab}$. They describe gravitational waves and have no
Newtonian analogue. Vector and scalar perturbations, on the other
hand, have Newtonian counterparts. Vector modes correspond to
rotational perturbations of the velocity field, while scalar modes
are associated with longitudinal density fluctuations. In these
lectures we will only consider the latter type of perturbations.

\subsection{The gauge problem}
It has long been known (see Lifshitz (1946)~\cite{L}) that the
study of cosmological perturbations is plagued by the so called
gauge problem, which stems from the fact that in perturbation
theory we actually deal with two spacetimes.\footnote{For a
extensive discussion on the gauge problem in cosmology the
interested reader is referred to articles by Sachs
(1966)~\cite{S}, Bardeen (1980)~\cite{Ba}, Ellis \& Bruni
(1989)~\cite{EB}, Stewart (1990), Bruni et al (1992) ~\cite{S}.}
The physical spacetime, ${\cal W}$, corresponding to the real
universe and a fictitious one, $\overline{\cal W}$, which defines
the unperturbed background. The latter is usually represented by
the homogeneous and isotropic FRW models. To proceed we need to
establish an one-to-one correspondence $\varphi:\overline{\cal
W}\mapsto{\cal W}$, namely a gauge, between the two spacetimes.
When a coordinate system is introduced in $\overline{\cal W}$ the
gauge caries it to ${\cal W}$ and vice versa, thus defining a
background spacetime into the real universe. In practice what a
gauge does is define the ``slicing'' and ``threading'' of the
spacetime into spacelike hypersurfaces and timelike worldlines
respectively. Any change in $\varphi:\overline{\cal W}\mapsto{\cal
W}$, keeping the background coordinates fixed, is known as a
``gauge transformation''. The latter induces a coordinate
transformation in the physical spacetime but also changes the
event in ${\cal W}$ that is associated with a given event in
$\overline{\cal W}$. Therefore, gauge transformations should be
distinguished from coordinate transformations which simply relabel
events.\footnote{In practice, we often talk about coordinate
changes in ${\cal W}$, since a frame in ${\cal W}$ corresponds,
via $\varphi:\overline{\cal W}\mapsto{\cal W}$, to one already
established in $\overline{\cal W}$. Therefore, the coordinate
choice in ${\cal W}$ also determines the gauge between
$\overline{\cal W}$ and ${\cal W}$. In this respect, gauge
transformations can be represented as coordinate changes in ${\cal
W}$.} The problems stem from our freedom to make gauge
transformations. By definition, a perturbation in a given quantity
is the difference between its value at some event in the real
spacetime and its value at the corresponding event in the
background. This means that even if a quantity behaves as a scalar
under coordinate changes, its perturbation will not be invariant
under gauge transformations, provided that the quantity in
question is non-zero and time dependent in the background. As a
result, one might end up with spurious gauge modes in the
solutions, which have no physical meaning whatsoever. Typical
example are density perturbations. Given that the density is a
time dependent scalar in the background, density perturbations are
not invariant under gauge transformations that change the
correspondence between the hypersurfaces of simultaneity in ${\cal
W}$ and $\overline{\cal W}$. Note that on scales bellow the
horizon the hypersurfaces of constant time are physically
unambiguous and the gauge choice is well motivated. On
super-horizon scales, however, one is free to choose between
gauges that give entirely different results for the time
dependence of the density perturbation. It becomes clear,
therefore, that Newtonian perturbations are not plagued by gauge
ambiguities but early universe studies are.

There are two ways of dealing with the gauge problem. The first,
which will be our approach, is to choose a particular gauge and
compute everything there. If the gauge choice is well motivated,
the perturbed variables will be easy to interpret. However, the
task of selecting the best gauge for any given situation,
something known as the ``fitting problem'' in cosmology, is not
always trivial. The second approach is to describe perturbations
using gauge-invariant variables. In a very influential paper,
Bardeen (1980)~\cite{Ba} introduced a fully gauge-invariant
approach following earlier work by Gerlach \& Sengupta (1978) (see
also Kodama \& Sasaki (1984) for an extensive review). Bardeen's
approach, however, is of considerable complexity as it determines
a set of gauge-invariant quantities that are related to density
perturbations but are not perturbations themselves. Building on
earlier work by Hawking (1966), Stewart \& Walker (1974) and Olson
(1976), Ellis \& Bruni (1989)~\cite{EB} formulated a fully
covariant gauge-invariant treatment of cosmological perturbations.
Their approach, which is of high mathematical elegance and
physical transparency, has the additional advantage of starting
from the fully non-linear equations before linearizing them about
a chosen background.

\subsection{The relativistic equations}
We proceed by adopting what is usually known as the covariant
Lagrangian approach to cosmology, which is also a direct extension
of the Newtonian treatment. The basic philosophy is to employ
locally defined quantities and derive their evolution equations
along the worldlines of the comoving observers. We start by
assuming a congruence of timelike worldlines tangent to the
4-velocity vector $u_a$. The later determines the motion of a
fundamental observer comoving with the fluid and is normalized so
that $u_0=1=u^{_0}$ and $u^{\alpha}=0$, with $\alpha=1,2,3$. Note
that throughout these lectures we have set $c=1$. If $g_{ab}$ is
the spacetime metric, with signature (+~-~-~-), then
$h_{ab}=g_{ab}-u_au_b$ is the metric of the three-dimensional
spaces orthogonal to the observer's motion (note that
$h_{ab}u^b=0$), which define the observer's instantaneous rest
space. Also, if $\nabla_a$ is the covariant derivative relative to
$g_{ab}$, then ${\rm D}_a=h_a{}^b\nabla_b$ is the covariant
derivative operator on these hypersurfaces provided that there is
no vorticity. Also, ${\rm D}^2=h^{ab}\nabla_a\nabla_b$ is the
associated Laplacian. Finally, an overdot indicates
differentiation along $u_a$, namely derivatives with respect to
proper time $\tau$. For example, $\dot{u}_a=u^b\nabla_bu_a$ is the
4-acceleration. The above procedure determines our gauge choice.

The basis of the relativistic analysis is the Einstein Field
Equations (EFE) describing the interaction between matter and
spacetime geometry\footnote{For an updated and extensive
discussion on relativistic cosmological models the reader is
referred to Ellis \& van Elst (1999)~\cite{EvE}.}
\begin{equation}
G_{ab}\equiv R_{ab}- {\textstyle{1\over2}}Rg_{ab}= \kappa T_{ab}-
\Lambda g_{ab}\,,  \label{EFE}
\end{equation}
where $G_{ab}$ is the Einstein tensor, $R_{ab}$ is the Ricci
tensor, $R=R_a{}^a$ is the Ricci scalar, $T_{ab}$ is the total
energy-momentum tensor of the mater fields, $\Lambda$ is the
cosmological constant and $\kappa=8\pi G$. The Einstein tensor has
the extremely important property of having an identically
vanishing divergence, that is $\nabla^bG_{ab}=0$. When applied to
Eq.~(\ref{EFE}), the latter yields the conservation law
\begin{equation}
\nabla^bT_{ab}=0\,.  \label{cl}
\end{equation}
For a perfect fluid the stress-energy tensor takes the simple form
\begin{equation}
T_{ab}=\rho u_au_b+ ph_{ab}\,,  \label{Tab}
\end{equation}
where $\rho$ and $p$ are respectively the energy density and
pressure of the fluid. Here, similarly to the Newtonian analysis,
we assume a barotropic fluid with $p=p(\rho)$. Note that the
assumption of an isotropic pressure is not actually valid before
matter-radiation equality owing to particle diffusion and free
streaming effects (see Secs.~4.5.1 and 5.4). The relativistic
analogues of the continuity and Euler equations are obtained from
the timelike and spacelike parts of the conservation law
(\ref{cl}). In particular, substituting (\ref{Tab}) into
Eq.~(\ref{cl}) and then projecting along the observer's motion
(i.e.~contracting with $u_a$) we obtain the energy density
conservation law
\begin{equation}
\dot{\rho}+ 3H(\rho+p)=0\,,  \label{GRce}
\end{equation}
where $3H=\nabla^au_a={\rm D}^au_a$. On the other hand, by
projecting orthogonal to $u_a$ we arrive at the momentum density
conservation equation
\begin{equation}
(\rho+p)\dot{u}_a+ {\rm D}_ap=0\,. \label{GRmc}
\end{equation}
Equations (\ref{GRce}), (\ref{GRmc}) are supplemented by
\begin{equation}
R_{ab}u^au^b={\textstyle{1\over2}}\kappa(\rho+3p)\,.  \label{EFE1}
\end{equation}
The above relates the spaccetime geometry to the matter sources
and is the relativistic analogue of the Poisson equation. Note
that throughout these notes the cosmological constant has been set
to zero.

\subsection{The linear regime}
Before perturbing Eqs.~(\ref{GRce}), (\ref{GRmc}), we need to make
one additional step, which will fix our gauge completely. So far
we have not assigned time labels to the comoving hypersurfaces.
The proper time interval $\tau$ is position dependent on these
surfaces and does not provide an appropriate label for coordinate
time. Assuming that $t$ is a valid ordering label, one can show
(see e.g.~Padmanabhan (1993), Liddle \& Lyth (2000)~\cite{KT})
that
\begin{equation}
\frac{{\rm d}\tau}{{\rm d}t}=1-\frac{\delta p}{\rho+p}\,.
\label{tl}
\end{equation}
On using the above, the perturbed continuity equation gives
\begin{equation}
\left(\delta\rho\right)'=-3(\rho_0+p_0)\delta H-3H_0\delta\rho
\label{GRdotdelro}
\end{equation}
to first order, where the dash indicates differentiation with
respect to $t$. Defining $\delta=\delta\rho/\rho_0$ the above is
recast into
\begin{equation}
\delta'- 3wH_0{\delta}+ 3(1+w)\delta H=0\,, \label{GRdotdel}
\end{equation}
where $w=p_0/\rho_0$ determines the equation of state of the
medium. Also, starting from Eq.~(\ref{GRmc}) we have
\begin{equation}
\left(\delta H\right)'+ 2H_0\delta H+ {\textstyle{4\over3}}\pi
G\rho_0\delta +\frac{v_{\rm s}^2}{3(1+w)}{\rm D}^2\delta=0\,,
\label{GRdotvp}
\end{equation}
to first order, where $\delta H$ describes scalar deviations from
the smooth background expansion. We obtain Eq.~(\ref{GRdotvp}) by
taking the 4-divergence of Eq.~(\ref{GRmc}) and then employing the
Ricci identity. Applied to the 4-velocity vector the latter reads
$\nabla_{[a}\nabla_{b]}u_c=R_{abcd}u^d$, where $R_{abcd}$ is the
spacetime Riemman tensor ($R_{ab}=R^c{}_{acb}$ is the associated
Ricci tensor). To proceed further we need the following auxiliary
formulae
\begin{equation}
w'=3H_0(1+w)(w-v_{\rm s}^2) \hspace{5mm}{\rm and}\hspace{5mm}
H_0'=-{\textstyle{3\over2}}(1+w)H_0^2\,, \label{aux}
\end{equation}
where the latter is commonly referred to as the Raychaudhuri
equation. Taking the time derivative of Eq.~(\ref{GRdotvp}),
substituting (\ref{GRdotdel}), using the auxiliary relations
(\ref{aux}) and keeping up to first order terms we arrive at
\begin{equation}
\tilde{\delta}''_{({\rm k})}+\left(2-6w+3v_{\rm
s}^2\right)H_0\tilde{\delta}'_{({\rm k})}-
{\textstyle{3\over2}}\left(1+8w-3w^2-6v_{\rm
s}^2\right)H_0^2\tilde{\delta}_{({\rm k})}=-\frac{{\rm k}^2v_{\rm
s}^2}{a^2}\tilde{\delta}_{({\rm k})}\,, \label{GRddotdel}
\end{equation}
for the evolution of the ${\rm k}$-th perturbative mode. In
deriving the above we have also employed the Fourier decomposition
$\delta=\sum_{\rm k}\tilde{\delta}_{({\rm k})}Q^{({\rm k})}$ for
the density perturbation, with ${\rm D}_a\tilde{\delta}_{({\rm
k})}=0$. Here $Q_{({\rm k})}$ are scalar harmonics, with
$Q'_{({\rm k})}=0$, which are solutions of the Laplace-Beltrami
equation
\begin{equation}
{\rm D}^2Q_{({\rm k})}=-\frac{{\rm k}^2}{a^2}Q_{({\rm k})}\,.
\label{LB}
\end{equation}
The above immediately implies that ${\rm D}^2\delta=-({\rm
k}^2/a^2)\delta$, which explains the term in the right-hand side
of Eq.~(\ref{GRddotdel}). This equation may be thought of as the
relativistic counterpart of the Newtonian formula (\ref{Ndelk}).
One recovers the Newtonian limit by setting $w=0=v_{\rm s}^2$ in
the left hand side of (\ref{GRddotdel}), and using the background
Friedmann equation $H_0^2=8\pi G\rho_0/3$.

Note that in many applications it helps to recast
Eq.~(\ref{GRddotdel}) in terms of the scale factor. Introducing
$a$ as the independent variable we find
\begin{equation}
a^2\frac{d^2\tilde{\delta}_{({\rm k})}}{da^2}+
{\textstyle{3\over2}}(1-5w+2v_{\rm
s}^2)a\frac{d\tilde{\delta}_{({\rm k})}}{da}-
{\textstyle{3\over2}}\left[(1+8w-3w^2-6v_{\rm s}^2)
-{\textstyle{2\over3}}\frac{{\rm k}^2v_{\rm
s}^2}{a^2H_0^2}\right]\tilde{\delta}_{({\rm k})}=0\,,
\label{GRddotdel1}
\end{equation}
where we have employed the transformation laws ${\rm d}/{\rm
d}t=aH\,{\rm d}/{\rm d}a$ and ${\rm d}^2/{\rm d}t^2=a^2H^2\,{\rm
d}^2/{\rm d}a^2-[(1+3w)aH^2/2]\,{\rm d}/{\rm d}a$.

\subsection{Solutions}
We will now seek solutions to the relativistic perturbation
equations to supplement the Newtonian results of the previous
section. The cases to be considered are: (i)~Super-horizon sized
perturbations in the dominant non-relativistic component after
equilibrium. (ii)~Fluctuations in the relativistic matter before
matter-radiation equality both inside and outside the Hubble
radius.

\subsubsection{Perturbed Einstein - de Sitter universe}
The Newtonian analysis is valid for modes well within the Hubble
radius. On scales beyond $\lambda_{H}$, however, one needs to
engage general relativity even when dealing with non-relativistic
matter. For pressureless dust $w=0=v_{\rm s}^2$ and
Eq.~(\ref{GRddotdel1}) reads
\begin{equation}
a^2\frac{{\rm d}^2\delta}{{\rm d}a^2}+
{\textstyle{3\over2}}a\frac{{\rm d}\delta}{{\rm d}a}-
{\textstyle{3\over2}}\left(1- {\textstyle{2\over3}}\frac{{\rm
k}^2v_{\rm s}^2}{a^2H_0^2}\right)\delta=0\,, \label{GRESddotdel}
\end{equation}
where again we have dropped the tilde and the wavenumber for
simplicity. For modes lying beyond the Hubble radius
$\lambda\gg\lambda_{H}$ and ${\rm k}^2v_{\rm s}^2/a^2H_0^2\ll1$.
On these scales the above reduces to
\begin{equation}
a^2\frac{{\rm d}^2\delta}{{\rm d}a^2}+
{\textstyle{3\over2}}a\frac{{\rm d}\delta}{{\rm d}a}-
{\textstyle{3\over2}}\delta=0\,. \label{GRESddotdel1}
\end{equation}
with
\begin{equation}
\delta={\cal C}_1a+{\cal C}_2a^{-3/2}\,. \label{GRES}
\end{equation}
Thus, after recombination large-scale perturbations in the
non-relativistic component grow as $\delta_{\rm b}\propto a\propto
t^{2/3}$.

\subsubsection{The radiation dominated era}
Before equipartition radiation dominates the energy density of the
universe and $w=1/3=v_{\rm s}^2$. During this period
Eq.~(\ref{GRddotdel1}) gives
\begin{equation}
a^2\frac{d^2\delta_{\gamma}}{da^2}- 2\left(1
-{\textstyle{1\over6}}\frac{{\rm
k}^2}{a^2H_0^2}\right)\delta_{\gamma}=0\,. \label{GRrddotdel}
\end{equation}
On large scales, when $\lambda\gg\lambda_{H}$ and ${\rm
k}^2/a^2H_0^2\ll1$, the above reduces to
\begin{equation}
a^2\frac{d^2\delta_{\gamma}}{da^2}- 2\delta_{\gamma}=0\,,
\label{GRrlsddotdel}
\end{equation}
with a power law solution of the form
\begin{equation}
\delta_{\gamma}={\cal C}_1a^2+ {\cal C}_2a^{-1}\,,  \label{GRrls}
\end{equation}
where ${\cal C}_1$, ${\cal C}_2$ are constants. Hence, before
matter-radiation equality large-scale perturbations in the
radiative fluid grow as $\delta_{\gamma}\propto a^{1/2}$. Note
that Eq.~(\ref{GRrlsddotdel}) also governs the evolution of the
non-relativistic component (baryonic or not), since it does not
incorporate any pressure effects. Therefore, solution
(\ref{GRrls}) also applies to baryons and collisionless matter.

On sub-horizon scales, with $\lambda\ll\lambda_{H}$ and ${\rm
k}^2/a^2H_0^2\gg1$ Eq.~(\ref{GRrddotdel}) becomes
\begin{equation}
a^2\frac{d^2\delta_{\gamma}}{da^2}+
{\textstyle{1\over3}}\frac{{\rm
k}^2}{a^2H_0^2}\delta_{\gamma}=0\,. \label{GRrssddotdel}
\end{equation}
and admits the oscillatory solution
\begin{equation}
\delta_{\gamma}={\cal C}{\rm e}^{\,i\lambda_{H}/\lambda}\,,
\label{GRrss}
\end{equation}
where ${\cal C}=$ constant. Thus, in the radiation era small-scale
fluctuations in the relativistic component oscillate like sound
waves. Given that $\lambda_{H}/\lambda\gg1$, the oscillation
frequency is very high. As a result,
$\langle\delta_{\gamma}\rangle\simeq0$ on scales well bellow the
Hubble radius. In other words, the radiative fluid is expected to
have a smooth distribution on small scales (see Sec.~2.6.2).

Note that in the radiation era the transition from growing to
oscillatory modes occurs at $\lambda\sim\lambda_{H}$, which
implies that before equipartition the role of the Jeans length is
played by the Hubble radius.

\subsection{Summary}
General relativity is necessary on scales outside the horizon and
also when studying perturbations in the relativistic component.
Prior to equilibrium, fluctuations in the radiative fluid grow as
$\delta_{\gamma}\propto a^2\propto t$ for wavelengths larger the
Hubble radius. Super-horizon sized matter perturbations also grow
at the same rate. On small scales, however, fluctuations in the
photon density oscillate rapidly. As a result, the small-scale
distribution of the radiative fluid, and of the tightly coupled
baryons, remains smooth. After equipartition perturbations in the
non-relativistic matter grow proportionally to the scale factor.

\section{Baryonic structure formation}
We will now consider cosmological models where baryons are the
dominant form of matter. It should be made clear at the outset,
however, that purely baryonic models cannot successfully explain
the origin of the observed structure. Nevertheless, it is
important to look at the details of these scenarios. After all,
whatever the dominant form of matter, baryons do exist in the
universe and it is crucial to study their behavior. The key issue
is understanding the interaction between baryonic matter and
radiation during the plasma epoch. The simplest way of doing so is
by looking at models containing these two components only.

\subsection{Adiabatic and isothermal perturbations}
Before recombination the universe was a mixture of ionized matter
and radiation interacting via Thomson scattering.\footnote{For
simplicity we will neglect the presence of Helium nuclei and the
role of the neutrinos.} When dealing with the pre-recombination
plasma we distinguish between two types of perturbations, namely
between ``adiabatic" and ``isothermal" disturbances (Zeldovich
(1967)~\cite{Z}). The former include fluctuations in both the
radiation and the matter component (i.e.~$\delta_{\rm
b}\,,\delta_{\gamma}\neq0$), whereas in the latter only the matter
component is perturbed (i.e.~$\delta_{\rm b}\neq0$ but
$\delta_{\gamma}=0$). Before recombination, a generic perturbation
can be decomposed into a superposition of independently
propagating adiabatic and isothermal modes. After matter and
radiation have decoupled, however, perturbations evolve in the
same way regardless of their original nature. Because there is no
interaction between baryons and photons and radiation field is
dynamically negligible, the universe behaves as a single-fluid
dust model.

\subsubsection{Adiabatic perturbations}
Adiabatic (or isentropic) modes contain fluctuations both in the
matter and the radiation components, while keeping the entropy per
baryon conserved. Note that if ${\cal S}=4\rho_{\gamma}/3{\rm
T}_{\gamma}$ is the photon entropy, then ${\rm S}={\cal S}/k_{\rm
B}{\rm n}_{\rm b}$ is the entropy per baryon, where ${\rm
T}_{\gamma}$ is the photon temperature, $k_{\rm B}$ is the
Boltzmann constant and ${\rm n}_{\rm b}$ is the baryon number
density. Given that $\rho_{\gamma}\propto{\rm T}_{\gamma}^4$ and
$\rho_{\rm b}\propto{\rm n}_{\rm b}$, the entropy per baryon
satisfies the relation ${\rm
S}\propto\rho_{\gamma}^{3/4}/\rho_{\rm b}$. Consequently one
arrives at
\begin{equation}
\frac{\delta{\rm S}}{\rm
S}=0\,\Rightarrow\,\,\,\left({\textstyle{3\over4}}\frac{\delta\rho_{\gamma}}{\rho_{\gamma}}
-\frac{\delta\rho_{\rm b}}{\rho_{\rm
b}}\right)=0\,\Rightarrow\,\,\,\delta_{\rm
b}={\textstyle{3\over4}}\delta_{\gamma}\,, \label{adm}
\end{equation}
which is known as the condition for adiabaticity. A set of density
contrasts satisfying the above requirement consists an adiabatic
perturbation. The later are naturally generated in the simplest
inflationary models through the vacuum fluctuation of the inflaton
field (see Liddle \& Lyth (2000)~\cite{KT}).

\subsubsection{Isothermal and isocurvature perturbations}
Isothermal modes include only matter fluctuations, while the
radiation field is assumed to be uniformly distributed. This means
that the radiation temperature is also uniform (recall that
$\rho_{\gamma}\propto{\rm T}_{\gamma}^4$), which explains the name
isothermal. This type of perturbation is closely related to the
``isocurvature'' modes, where $\delta{_{\rm
b}}\,\delta_{\gamma}\neq0$ but $\delta\rho=0$ ($\rho=\rho_{\rm
b}+\rho_{\gamma}$). This implies that the geometry of the
3-dimensional spatial hypersurfaces remains unaffected, hence the
name isocurvature. Note that isocurvature perturbations can still
affect, via the pressure, the 4-dimensional spacetime geometry.
For such modes we have
\begin{equation}
\delta\rho=0\,\Rightarrow\,\,\,\rho_{\rm b}\delta_{\rm b}+
\rho_{\gamma}\delta_{\gamma}=0\,\Rightarrow\,\,\,\frac{\delta_{\gamma}}{\delta_{\rm
b}}=-\frac{\rho_{\rm b}}{\rho_{\gamma}}\,.  \label{iscm}
\end{equation}
Clearly, when $\rho_{\gamma}\gg\rho_{\rm b}$, as it happens in the
radiation era, one finds that $\delta_{\gamma}\ll\delta_{\rm b}$,
which explains why isocurvature modes are very often referred to
as isothermal. Unlike adiabatic disturbances, isocurvature
perturbations are usually absent from the simplest models of
inflation. They can still be produced, however, in multi component
inflationary scenarios by the vacuum fluctuation of a field other
than the inflaton (see Liddle \& Lyth (2000)~\cite{KT}). As stated
earlier, the distinction between adiabatic and isothermal
fluctuations is meaningful only prior to recombination, when
matter and radiation are tightly coupled. After decoupling,
perturbations in the matter component evolve as if they were
effectively isothermal.

\subsection{Evolution of the sound speed}
The different nature of the adiabatic from the isothermal
perturbations means that each type of disturbance has its own
discrete signatures. The key issue is the evolution of the sound
speed, since it determines the scale of gravitational instability.
Here, we consider the evolution of the sound speed for the
adiabatic and the isothermal modes during the plasma era.

\subsubsection{The adiabatic sound speed}
In a mixture of radiation and matter the total density and
pressure are $\rho=\rho_{\gamma}+\rho_{\rm b}$ and $p\simeq
p_{\gamma}=\rho_{\gamma}/3$ respectively (recall that
$p_{\gamma}=\rho_{\gamma}/3$ and $p_{\rm b}\simeq0$). Hence, the
adiabatic sound speed is given by
\begin{equation}
v_{\rm s}^{({\rm a})}= \left(\frac{\partial p}{\partial
\rho}\right)^{1/2}\simeq
{\textstyle{1\over\sqrt{3}}}\left(1+\frac{3\rho_{\rm
b}}{4\rho_{\gamma}}\right)^{-{1/2}}\,, \label{ass}
\end{equation}
where we have used the adiabatic condition $\partial\rho_{\rm
b}/\partial\rho_{\gamma}=3\rho_{\rm b}/4\rho_{\gamma}$ (see
Eq.~(\ref{adm})). In the radiation era $\rho_{\gamma}\gg\rho_{\rm
b}$ ensuring that $v_{\rm s}^{({\rm a})}\simeq1/\sqrt{3}$. In the
interval between equipartition and decoupling, when $\rho_{\rm
b}\gg\rho_{\gamma}$, Eq.~(\ref{ass}) gives $v_{\rm s}^{({\rm
a})}\simeq\sqrt{4\rho_{\gamma}/3\rho_{\rm b}}\propto a^{-1/2}$. In
particular, employing the relations $\rho_{\gamma}=\rho_{\rm
eq}[(1+z)/(1+z_{\rm eq})]^4$ and $\rho_{\rm b}=\rho_{\rm
eq}[(1+z)/(1+z_{\rm eq})]^3$, we find (see Coles \& Lucchin
(1995)~\cite{KT})
\begin{equation}
v_{\rm s}^{({\rm a})}\simeq2\times10^8\left(\frac{1+z}{1+z_{\rm
eq}}\right)^{1/2}\,{\rm cm}/{\rm sec}\,. \label{adss}
\end{equation}
Note that throughout these notes we assume that $z_{\rm
eq}>z_{_{\rm rec}}$, that is equipartition occurs prior to
recombination. Given that $1+z_{\rm
eq}\simeq4\times10^4(\Omega_{\rm b}h^2)$ and $1+z_{\rm
rec}\simeq10^3$, our assumption holds only if $\Omega_{\rm
b}h^2\geq0.02$.

\subsubsection{The isothermal sound speed}
The sound speed associated with isothermal fluctuations is that of
a monatomic gas
\begin{equation}
v_{\rm s}^{({\rm i})}=\left(\frac{\partial p_{\rm
b}}{\partial\rho_{\rm b}}\right)^{1/2}= \left(\frac{\gamma\,k_{\rm
B}{\rm T}_{\rm b}}{3m_{\rm p}}\right)^{1/2}\,, \label{iss}
\end{equation}
where ${\rm T}_{\rm b}$ is the matter temperature, $m_{\rm p}$ is
the proton mass and $\gamma=5/3$ for hydrogen. Before decoupling
photons and baryons are still tightly coupled and share the same
temperature (i.e.~${\rm T}_b\simeq{\rm T}_{\gamma}$). Equation
(\ref{iss}) then implies that $v_{\rm s}^{({\rm i})}\propto{\rm
T}_{\gamma}^{1/2}\propto a^{-1/2}$. In particular, given that
$(T_{\gamma})_{\rm rec}\simeq4\times10^3$ and using the relation
${\rm T}_{\gamma}=({\rm T}_{\gamma})_{\rm rec}(1+z)/(1+z_{\rm
rec})$, we obtain (see Coles \& Lucchin (1995)~\cite{KT})
\begin{equation}
v_{\rm s}^{({\rm i})}=5\times10^5\left(\frac{1+z}{1+z_{\rm
eq}}\right)^{1/2}\,{\rm cm}/{\rm sec}\,, \label{iss1}
\end{equation}
for $z<z_{\rm rec}$. After decoupling the isothermal sound speed
obeys Eq.~(\ref{iss1}) as long as ${\rm T}_{\rm b}\simeq
T_{\gamma}$ (i.e.~for $z>300$). Subsequently to that and until the
time of reheating ${\rm T}_{\rm b}\neq{\rm T}_{\gamma}$ and
$v_{\rm s}\propto{\rm T}_{\rm b}^{1/2}\propto a^{-1}$.

As it was pointed out earlier, in the post-recombination we can no
longer distinguish between adiabatic and isothermal perturbations.
After decoupling the sound speed of matter perturbations coincides
with the isothermal one. Thus, at recombination the sound speed of
the adiabatic disturbances drops from $v_{\rm s}^{({\rm
a})}\propto(p_{\gamma}/\rho_{\rm b})^{1/2}$ to $v_{\rm s}^{({\rm
a})}\propto(p_{\rm b}/\rho_{\rm b})^{1/2}$ (see Eqs.~(\ref{ass}),
(\ref{iss})). Given that $p_{\gamma}\sim{\rm n}_{\gamma}{\rm
T}_{\gamma}$ and $p_{\rm b}\sim{\rm n}_{\rm b}{\rm T}_{\rm b}$
with ${\rm n}_{\gamma}/{\rm n}_{\rm b}\simeq10^8$, the reduction
in $v_{\rm s}^{({\rm a})}$ is very large (recall that ${\rm
T}_{\rm b}\simeq{\rm T}_{\gamma}$ through recombination).
Following such a drop in the adiabatic sound speed at decoupling
suggests, one anticipates a similarly dramatic decrease in the
Jeans length at around the same time.

\subsection{Evolution of the Jeans length and the Jeans mass}
\subsubsection{Adiabatic modes}
The difference in the sound speed between adiabatic and isothermal
fluctuations means that their respective Jeans length and Jeans
mass also differ. This in turn implies that adiabatic and
isothermal modes become gravitationally unstable at different
scales. In particular, using definition (\ref{Jl}) and the results
of Sec.~4.2, we find that before recombination the adiabatic Jeans
length evolves as
\begin{equation}
\lambda_{\rm J}^{({\rm a})}\propto\left\{\begin{array}{l}
a^2\hspace{10mm}z>z_{\rm eq}\,,\\ a\hspace{12mm}z_{\rm
eq}>z>z_{\rm rec}\,.\\\end{array}\right.  \label{aJl}
\end{equation}
In deriving the above we have taken into account that $v_{\rm
s}^{({\rm a})}\simeq\sqrt{1/3}$ and
$\rho\simeq\rho_{\gamma}\propto a^{-4}$ for $z>z_{\rm eq}$, while
$v_{\rm s}^{({\rm a})}\propto a^{-1/2}$ and $\rho\simeq\rho_{\rm
b}\propto a^{-3}$ when $z_{\rm eq}>z>z_{\rm rec}$. On using
definition (\ref{Jm}), the above result translates into
\begin{equation}
M_{\rm J}^{({\rm a})}\propto\left\{\begin{array}{l}
a^3\hspace{22mm}z>z_{\rm eq}\,,\\ {\rm
constant}\hspace{12mm}z_{\rm eq}>z>z_{\rm rec}\,,\\
\end{array}\right.  \label{aJm}
\end{equation}
with $(M_{\rm J}^{({\rm a})})_{\rm rec}\simeq3\times10^{15}(\Omega
h^2)^{-2}~{\rm M}_{\odot}$ (see Coles \& Lucchin
(1995)~\cite{KT}).\footnote{For consistency the values for the
baryonic Jeans mass given in this section have all been quoted
from Coles \& Lucchin~\cite{KT}, with the assumption that
$\Omega=\Omega_{\rm b}=1$.} Consequently, in the adiabatic
scenario, the first scales to become gravitationally unstable and
collapse soon after decoupling have the size of a supercluster of
galaxies.

\subsubsection{Isothermal modes}
Throughout the plasma era ${\rm T}_{\rm b}\simeq{\rm T}_{\gamma}$
and the isothermal sound speed evolves as $v_{\rm s}^{({\rm
i})}\propto a^{-1/2}$, which substituted into definition
(\ref{Jl}) gives
\begin{equation}
\lambda_{\rm J}^{({\rm i})}\propto\left\{\begin{array}{l}
a^{3/2}\hspace{7.5mm}z>z_{\rm eq}\,,\\
a\hspace{12mm}z_{\rm eq}>z>z_{\rm rec}\,.\\
\end{array}\right.  \label{iJl}
\end{equation}
The corresponding Jeans mass is governed by
\begin{equation}
M_{\rm J}^{({\rm i})}\propto\left\{\begin{array}{l}
a^{3/2}\hspace{19mm}z>z_{\rm eq}\,,\\ {\rm
constant}\hspace{12mm}z_{\rm eq}>z>z_{\rm rec}\,.\\
\end{array}\right.  \label{iJm}
\end{equation}
with $(M_{\rm J}^{({\rm i})})_{\rm rec}\sim5\times10^4(\Omega
h^2)^{-1/2}~{\rm M}_{\odot}$ (see Coles \& Lucchin
(1995)~\cite{KT}). In the isothermal models the first sizes to
collapse are of the order of a globular star cluster. According to
results (\ref{aJm}), (\ref{iJm}), the Jeans mass, of both
adiabatic and isothermal perturbations, increases during the
radiation era but it remains constant in the interval between
equipartition and recombination. So, $M_{\rm J}$ takes its maximum
possible value in models with $z_{\rm eq}=z_{\rm rec}$. Note the
huge drop, of the order of $10^{11}~{\rm M}_{\odot}$, in $M_{\rm
J}^{(a)}$ around decoupling. This is the result of a similarly
large drop of the adiabatic sound speed at the same time (see
previous section).

After recombination the Jeans length and the Jeans mass of matter
perturbations are taken to coincide with $\lambda_{\rm J}^{({\rm
i})}$ and $M_{\rm J}^{({\rm i})}$ respectively. The latter evolves
as $M_{\rm J}\propto a^{-3/2}$, since $v_{\rm s}^{({\rm
i})}\propto a^{-1}$ for $z<z_{\rm rec}$.

\subsection{Evolution of the Hubble mass}
An additional important scale for structure formation is that of
the Hubble mass $M_{H}$ defined as the total amount of mass
contained within a sphere of radius $\lambda_{H}/2$,
\begin{equation}
M_{H}={\textstyle{4\over3}}\pi\rho\left(\frac{\lambda_{H}}{2}\right)^3\,,
\label{Hm}
\end{equation}
where $\rho$ is the density of the perturbed component and
$\lambda_{H}$ is the Hubble radius. In these lectures we will only
consider the baryonic Hubble mass (i.e.~$\rho=\rho_{\rm b}$
always). Also, recall that we will only be dealing with models
where the Hubble radius is effectively identical to the particle
horizon. In this respect, $\lambda_{H}$ and $M_{H}$ define the
scale over which the different parts of a perturbation are in
causal contact. Note that a mass scale $M$ is said to be entering
the Hubble radius when $M=M_{H}$. Given that $\lambda_{H}\propto
t\propto a^2$ before equilibrium and $\lambda_{H}\propto t\propto
a^{3/2}$ for $t>t_{\rm eq}$ we obtain the evolution law
\begin{equation}
M_{H}\propto\left\{\begin{array}{l}
a^3\hspace{15mm}z>z_{\rm eq}\,,\\
a^{3/2}\hspace{12mm}z_{\rm eq}>z\,.\\
\end{array}\right.  \label{Hm1}
\end{equation}
It should be emphasized that before matter-radiation equality the
baryonic Jeans mass is of the same order with the Hubble mass.
Indeed, when radiation dominates $\rho\simeq\rho_{\gamma}$, which
means that $\lambda_{\rm J}\sim(G\rho_{\gamma})^{-1/2}$ and
$\lambda_{H}\sim(G\rho_{\gamma})^{-1/2}$. Given that and using
definitions (\ref{Jm}) and (\ref{Hm}) one can easily verify that
$M_{\rm J}\simeq M_{H}$ throughout the radiation epoch.

\subsection{Dissipative effects}
To this point we have treated the cosmic medium purely
gravitationally, as a perfect fluid, ignoring any dissipative
effects. In the process we have established two key physical
scales, the Jeans mass and the Hubble mass, which play an
important role in all structure formation models. We now turn our
attention to other physical processes that can modify the purely
gravitational evolution of perturbations. In baryonic models the
most important physical phenomenon is the interaction between
baryons and photons in the pre-recombination era, and the
consequent dissipation due to viscosity and heat conduction.

\subsubsection{Collisional damping of adiabatic perturbations}
Adiabatic perturbations in the photon-baryon plasma suffer from
collisional damping around the time of recombination because the
perfect fluid approximation breaks down. As we approach
decoupling, the photon mean free path increases and photons can
diffuse from the overdense into the underdense regions, thereby
smoothing out any inhomogeneities in the primordial plasma. The
effect is known as collisional dissipation or ``Silk damping''
(Silk (1967)~\cite{Si}). A detailed treatment requires integrating
the Boltzmann equation through recombination. Here we will only
obtain an estimate of the effect. To begin with, we consider the
physical (proper) distance associated with the photon mean free
path
\begin{equation}
\ell_{\gamma}=\frac{1}{X_{\rm e}{\rm n}_{\rm e}\sigma_{\rm
T}}\simeq 10^{29}a^3X_{\rm e}^{-1}\left(\Omega_{\rm
b}h^2\right)^{-1}\,{\rm cm}\,,  \label{pmp}
\end{equation}
where $X_{\rm e}$ is the electron ionization factor, ${\rm n}_{\rm
e}\propto a^{-3}$ is the number density of the free electrons and
$\sigma_{\rm T}$ is the cross section for Thomson scattering.
Clearly, all baryonic perturbations with wavelengths smaller than
$\ell_{\gamma}$ will be smoothed out by photon free streaming. The
perfect fluid assumption breaks down completely when
$\lambda\ll\ell_{\gamma}$. Damping, however, occurs on scales much
larger than $\ell_{\gamma}$ as the photons slowly diffuse from the
overdense into the underdense regions, dragging along the still
tightly coupled baryons. Within a time interval $\Delta t$ a
photon suffers ${\rm N}=\Delta t/\ell_{\gamma}$ collisions and
performs a random walk with mean square coordinate (i.e.~comoving)
displacement given by
\begin{equation}
\langle\Delta x\rangle^2=\,\,{\rm
N}\left(\frac{\ell_{\gamma}}{a}\right)^2=\,\,
\frac{\ell_{\gamma}}{a^2}\,\Delta t\,,  \label{mscd}
\end{equation}
where $(\ell_{\gamma}/a$) is the coordinate distance between
successive collisions. Integrating the above up to recombination
time we obtain the total coordinate distance travelled by a
typical photon
\begin{eqnarray}
x_{\rm S}^2&=&\int_0^{t_{\rm rec}}\frac{\ell_{\gamma}}{a^2}{\rm
d}t\,\,=\,\,\frac{(\ell_{\gamma})_{\rm rec}}{a_{\rm rec}^2t_{\rm
rec}^{2/3}}\int_0^{t_{\rm
rec}}t^{2/3}{\rm d}t \nonumber\\
&=&{\textstyle{3\over5}}\frac{(\ell_{\gamma})_{\rm rec}t_{\rm
rec}}{a_{\rm rec}^2}\,,  \label{clS}
\end{eqnarray}
on using the relations $\ell_{\gamma}=(\ell_{\gamma})_{\rm
rec}(a/a_{\rm rec})^3$ and $a=a_{\rm rec}(t/t_{\rm rec})^{2/3}$
for the photon mean free path and the scale factor of the universe
respectively. The latter is a reasonable approximation given that
earlier on the coupling between the photons and the electron was
too tight for the photons to move at all. The physical scale
associated with the above result is (see Kolb \& Turner
(1990)~\cite{KT})
\begin{equation}
\ell_{\rm S}= ax_{\rm S}=
\sqrt{{\textstyle{3\over5}}(\ell_{\gamma})_{\rm rec}t_{\rm
rec}}\,\,\simeq\,\,3.5\left(\Omega h^2\right)^{-3/4}\,{\rm Mpc}\,,
\label{lS}
\end{equation}
and the associated mass scale, which is known as the ``Silk
mass'', is given by\footnote{Given that $t_{\rm
rec}\gg(\ell_{\gamma})_{\rm rec}$ we have $(\ell_{\gamma})_{\rm
rec}t_{\rm rec}\gg(\ell_{\gamma})_{\rm rec}^2$, which ensures that
$\ell_{\rm S}\gg(\ell_{\gamma})_{\rm rec}$. Consequently, Silk
damping is felt on scales much larger than the mean free path of a
typical photon.}
\begin{equation}
M_{\rm S}={\textstyle{4\over3}}\pi\rho_{\rm
b}\left(\frac{\ell_{\rm
S}}{2}\right)^3\simeq6.2\times10^{12}\left(\Omega
h^2\right)^{-5/4}\,{\rm M}_{\odot}\,, \label{Sm}
\end{equation}
for adiabatic baryonic perturbations. Note that if we had not
included dissipation, the amplitude of an acoustic wave on a mass
scale smaller than the Jeans mass would have remained constant
during the radiation era and then decayed as $t^{-1/6}$ in the
interval between equilibrium and decoupling (see
Eq.~(\ref{EdS2})). The dissipative process we considered above
causes the amplitude of these waves to decrease at a rate that
depends on the size of the perturbation. The final result is that
fluctuations on scales bellow the Silk mass are completely
obliterated by the time of recombination and no structure can form
on these scales. Alternatively, one might say that adiabatic
perturbations have very little power left on small scales.

\subsubsection{Freezing-in of isothermal perturbations}
Consider isothermal perturbations in the pre-recombination era on
scales larger than the Jeans mass. As the matter particles try to
move around, they encounter a viscous friction force due to
Thomson scattering with the smoothly distributed background
photons. This acts an effective drag force on the baryons causing
the isothermal mode to freeze-in throughout the plasma era.
Qualitatively, one can explain this effect on the basis of the
following simple physical argument. Consider the viscous force,
per unit mass, due to the aforementioned ``radiation drag''
\begin{equation}
F_{\rm T}\sim\frac{v}{t_{\rm T}}\sim\frac{\lambda}{t_{\rm T}t}\,,
\label{vf}
\end{equation}
where $\lambda$ is the wavelength of the perturbation and $t_{\rm
T}$ is the timescale for Thomson scattering. On the other hand,
the self-gravitating pull per unit mass of a baryonic perturbation
is
\begin{equation}
F_{G}\sim\frac{G\rho_{\rm b}V}{\lambda^2}\sim G\rho_{\rm
b}\lambda\sim\frac{\lambda}{t^2}\,, \label{gf}
\end{equation}
since $V\sim\lambda^3$ and $\rho_{\rm b}\sim1/Gt^2$. Before
recombination $t_{\rm T}\ll t$, implying that $F_G\ll F_{\rm T}$.
As a result isothermal perturbations cannot grow until matter and
radiation have decoupled. Note that the stagnation of isothermal
modes prior to recombination due to the radiation drag is not
related to the Meszaros effect discussed in Sec.~2.6.2. The latter
is purely kinematical and does not involve any interactions
between matter and radiation.

\subsection{Scenarios and problems}
Depending on the nature of the primeval perturbations one may
consider two scenarios of baryonic structure formation. Those
where the original inhomogeneities are of the adiabatic type and
those permeated by isothermal fluctuations. The adiabatic and
isothermal scenarios were in direct competition throughout the
1970s. One aspect of the confrontation was that the adiabatic
models were advocated by the Soviet school of astrophysicists led
by Zeldovich in Moscow, whereas the isothermal scenarios were
primarily an American affair promoted by Peebles and the Princeton
group. At the end, generic shortcomings in both models meant that
neither of these adversaries won the battle. So in the early
1980s, the purely baryonic models were sidelined by non-baryonic
dark matter scenarios (see Sec.~5).

\subsubsection{Adiabatic Scenarios}
Typically, adiabatic perturbations with sizes larger than the
maximum value of the Jeans mass, that is with $M>(M_{\rm J}^{({\rm
a})})_{\rm eq}\simeq3\times10^{15}(\Omega h^2)^{-2}~{\rm
M}_{\odot}$ experience uninterrupted growth. In particular they
grow as $\delta_{\rm b}\propto t$ before matter-radiation equality
and like $\delta_{\rm b}\propto t^{2/3}$ after equipartition.
Fluctuations on scales in the mass interval $(M_{\rm J}^{({\rm
a})})_{\rm eq}>M>(M_{\rm S})_{\rm rec}$ grow as $\delta_{\rm
b}\propto t$ while they are still outside the Hubble radius. After
horizon entry and until recombination these modes oscillate like
acoustic waves. The amplitude of the oscillation is constant
before equilibrium but decreases as $t^{-1/6}$ between
equipartition and recombination. After decoupling the modes become
unstable again and grow as $\delta_{\rm b}\propto t^{2/3}$.
Finally all perturbations on scales smaller than the value of the
Silk mass at recombination, that is with $M<(M_{\rm S})_{\rm
rec}\simeq6.2\times10^{12}\left(\Omega h^2\right)^{-5/4}\,{\rm
M}_{\odot}$ are eventually dissipated by photon diffusion. In
short, only fluctuations on scales exceeding that of a galaxy
cluster can survive the plasma epoch.
\begin{figure}
\epsffile{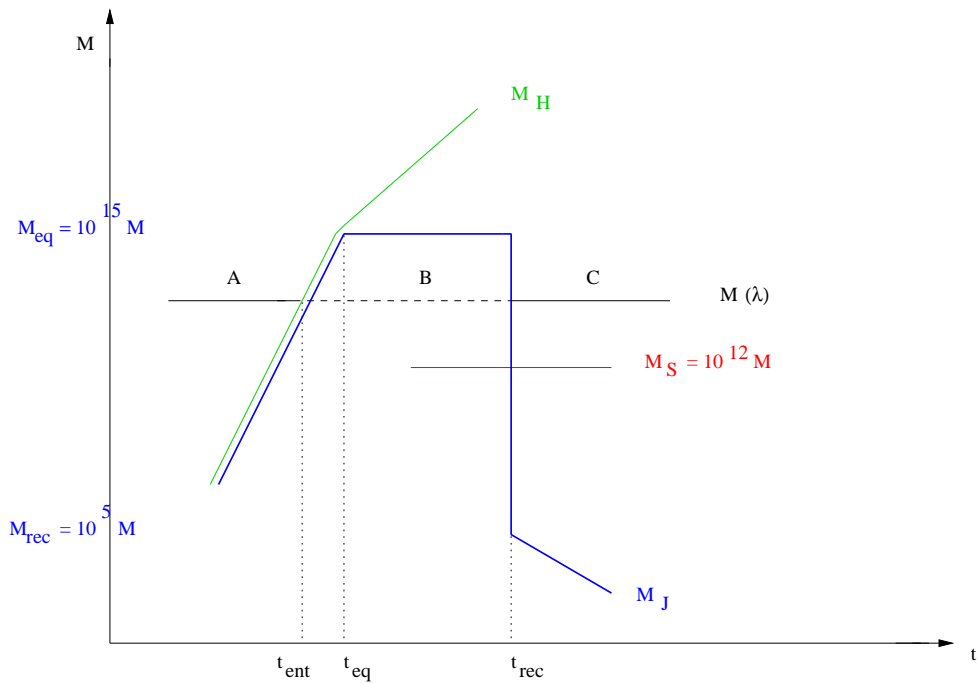}\caption{{\small Typical evolution of a
perturbed scale $M=M(\lambda)$ in adiabatic scenarios. During
stage A ($t<t_{\rm ent}<t_{\rm eq}$) the mode grows as
$\delta_{\rm b}\propto t$. Throughout stage B ($t_{\rm
ent}<t<t_{\rm eq}$) the perturbation oscillates. Finally, in stage
C ($t_{\rm rec}<t$) the mode becomes unstable again and grows as
$\delta_{\rm b}\propto t^{2/3}$ Note that fluctuations with size
smaller than $M_{\rm S}$ are wiped out by photon diffusion.¥}}
\label{fig: 1}
\end{figure}
Schematically, the evolution of an adiabatic mode with mass scale
$M(\lambda)$ is depicted in Figure~1. We distinguish between three
evolutionary stages A, B and C, depending on the size of the
perturbation and on the time of horizon crossing.

In adiabatic scenarios the smallest scales with any structure
imprint upon them at the time of recombination are as large as a
rich cluster of galaxies. Fluctuations on smaller scales have been
completely obliterated by Silk damping. After decoupling
perturbations grow steadily until their amplitude becomes of order
unity or larger. At that point the linear theory breaks down and
one needs to employ a different type of analysis. Qualitatively,
what happens is that those huge perturbations undergo anisotropic
collapse to form massive flattened objects known as ``pancakes''
or ``caustics'' (Zeldovich (1970)~\cite{Ze}). After pancake
formation, non-linear gas dynamics inside the collapsed structure
lead to shock generation and the subsequent cooling causes
fragmentation into smaller structures. Thus, in the adiabatic
scenario galaxies are born out of larger condensations in a
``top-down" fashion. The expected large-scale structure pattern is
one of large sheet-like filaments with enormous voids in between,
a picture that seems to fit qualitatively very well with
observations.

The main difficulty with the adiabatic scenario is that it
predicts angular fluctuations in the CMB temperature in excess
with observational limits. More specifically, in a $\Omega=1$
baryonic universe, one needs $\delta_{\rm b}\simeq10^{-3}$ at
recombination if structure were ever to form by today. In the
adiabatic picture, however, matter inhomogeneities are accompanied
by perturbations in the radiation field. This will inevitably lead
to temperature fluctuations of order $\delta{\rm T}/{\rm
T}\simeq\delta_{\gamma}\simeq10^{-3}$ at decoupling, which is in
direct disagreement with observations. To make things worse,
primordial nucleosynthesis restricts the baryonic contribution to
the total density of the universe down to $\Omega_{\rm
b}\simeq0.1$. Given that perturbations grow slower in open models
than in flat ones, a greater temperature fluctuation at
recombination is required if structure were to form by now.

\subsubsection{Isothermal scenarios}
In isothermal models structure formation proceeds very
differently. Isothermal fluctuations remain frozen-in throughout
the radiation era due to the radiation drag. After recombination
modes larger than $(M_{\rm J}^{({\rm i})})_{\rm
rec}\simeq5\times10^5(\Omega h^2)^{-1/2}~{\rm M}_{\odot}$ are the
first to collapse, while smaller scales are of no cosmological
relevance. Thus, in the isothermal picture the first masses to
condense out of the primordial plasma have the size of a globular
cluster. These first structures start clustering together via
gravitational instability to form successively larger
agglomerations. In other words, galaxies and galaxy clusters form
hierarchically in a ``bottom up'' fashion. Qualitatively, one
expects to see a roughly self-similar clustering pattern without
the dramatic structures of the adiabatic models.

Isothermal models do not have difficulties with the CMB
anisotropies, since by definition the radiation field is uniformly
distributed. In addition, the mass scale of the crucial first
generation of structures is too small. The major problems with
these scenarios is that the structure we see in the universe is
not purely hierarchical. Galaxies have their own individual
properties and galaxy clusters dot not just look like large
galaxies. In the isothermal models this individuality of galaxies
must be explained through some non-gravitational, probably
dissipative, process. In addition, isothermal disturbances seem
rather unnatural. Only very special physical processes can lead to
primordial fluctuations in the matter component leaving the
tightly coupled radiation field undisturbed. Particular
inflationary models, where the scalar field responsible for
generating the fluctuations is not the inflaton itself, may be
able to produce this special type of perturbations.

\section{Non-baryonic structure formation}
The difficulties faced by the purely baryonic models opened the
way for alternative structure formation scenarios, where the
dominant matter is non-baryonic. Further motivation for
considering models dominated by exotic matter species came from a
combination of observational data and theoretical prejudice.
Observations of the light element abundances, in particular,
require $\Omega_{\rm b}h^2<0.015$ to comply with primordial
nucleosynthesis calculations. On the other hand, dynamical
considerations seem to imply that $\Omega\simeq0.2$ and inflation
argues for $\Omega\simeq1$. If all these are true, our universe
must be dominated by exotic non-baryonic species to the extent
that the baryons are only a small fraction of the total matter.

\subsection{Non-baryonic cosmic relics}
An ongoing problem of the dark matter scenarios is that we still
do not exactly know neither the nature nor the masses of the
particles that make up the collisionless component of our
universe. High energy physics theories, however, provide us with a
whole ``zoo'' of candidates which are known as ``cosmic relics''
or ``relic WIMPs''. Typically, we distinguish between ``thermal''
and ``non-thermal'' relics. The former are kept in thermal
equilibrium with the rest of the universe until they decouple. A
characteristic example of this relic type is the massless
neutrino. Non-thermal relics on the other hand, such as axions,
magnetic monopoles and cosmic strings, have been out of
equilibrium throughout their lifetime. Thermal relics are further
subdivided into ``hot'' relics, which are still relativistic when
they decouple, and ``cold'' relics which go non-relativistic
before decoupling. A typical hot thermal relic is a light neutrino
with $m_{\nu}\simeq10~{\rm eV}$. The best motivated cold relic is
the lightest supersymmetric partner of the standard model
particles, with a predicted $\Omega_{WIMP}\sim1$. Note that
thermal relics with masses around $1$~keV are usually referred to
as ``warm dark matter''. Right-handed neutrinos, axinos and
gravitinos have all been suggested as potential warm relic
candidates.

\subsection{Evolution of the Jeans mass}
If the dark component is made up of weakly interacting species,
the particles do not feel each other's presence via collisions.
Each particle moves along a spacetime geodesic, while
perturbations modify these geodesic orbits. One can study the
response of the of the dark matter component by invoking an
effective pressure and treating the collisionless species as an
ideal fluid. The associated Jeans length is obtained similarly to
the baryonic one. When dealing with a collisionless species,
however, one needs to replace Eqs.~(\ref{Neqp1}), (\ref{Neqp2})
with the Liouville equation (see Coles \& Lucchin
(1995)~\cite{KT}). Then,
\begin{equation}
\lambda_{\rm J}=v_{\rm DM}\sqrt{\frac{\pi}{G\rho}}\,, \label{nbJl}
\end{equation}
where now $v_{\rm DM}$ is the velocity dispersion of the dark
matter component. The corresponding Jeans mass is
\begin{equation}
M_{\rm J}={\textstyle{4\over3}}\pi\rho_{\rm
DM}\left(\frac{\lambda_{\rm J}}{2}\right)^3\,,  \label{nbJm}
\end{equation}
where $\rho_{\rm DM}$ is the density of the non-baryonic matter.

\subsubsection{Hot thermal relics}
They decouple while they are still relativistic, that is $t_{\rm
dec}<t_{\rm nr}$, where we assume that $t_{\rm nr}<t_{\rm eq}$.
Throughout the relativistic regime $v_{\rm DM}\sim1$,
$\rho\simeq\rho_{\gamma}\propto a^{-4}$ and $\rho_{\rm DM}\propto
a^{-4}$, implying that $\lambda_{\rm J}^{({\rm h})}\propto a^2$
and $M_{\rm J}^{({\rm h})}\propto a^2$. Once the species have
become non-relativistic and until matter-radiation equality,
$v_{\rm DM}\propto a^{-1}$, $\rho\simeq\rho_{\gamma}\propto
a^{-4}$ and $\rho_{\rm DM}\propto a^{-3}$. Recall that the
particles have already decoupled, which means that ${\rm T}_{\rm
DM}\neq T_{\gamma}$. Consequently, $\lambda_{\rm J}^{({\rm
h})}\propto a$ and $M_{\rm J}^{({\rm h})}={\rm constant}$. After
equipartition $v_{\rm DM}\propto a^{-1}$ and $\rho\simeq\rho_{\rm
DM}\propto a^{-3}$, which translates into $\lambda_{\rm J}^{({\rm
h})}\propto a^{1/2}$ and $M_{\rm J}^{({\rm h})}\propto a^{-3/2}$.
Overall, the Jeans mass of hot thermal relics evolves as
\begin{equation}
M_{\rm J}^{({\rm h})}\propto\left\{\begin{array}{l}
a^2\hspace{20mm}z>z_{\rm nr}\,,\\ {\rm
constant}\hspace{10mm}z_{\rm nr}>z>z_{\rm eq}\,,\\
a^{-3/2}\hspace{15mm}z_{\rm eq}>z\,.
\end{array}\right.  \label{hJm}
\end{equation}
Clearly, $M_{\rm J}^{({\rm h})}$ reaches its maximum at $z_{\rm
nr}$. In fact, the highest possible value corresponds to particles
with $z_{\rm nr}=z_{\rm eq}$ such as neutrinos with
$m_{\nu}\simeq10~{\rm eV}$. In this case $(M_{\rm
J}^{({\nu})})_{\rm
max}\simeq3.5\times10^{15}(\Omega_{\nu}h^2)^{-2}~{\rm M}_{\odot}$
(see Coles \& Lucchin (1995)~\cite{KT}). For a typical hot thermal
relic $(M_{\rm J}^{({\rm h})})_{\rm max}\sim10^{12}-10^{14}~{\rm
M}_{\odot}$.

\subsubsection{Cold thermal relics}
Cold thermal relics decouple when they are already
non-relativistic (i.e.~$t_{\rm nr}<t_{\rm dec}<t_{\rm eq}$). Thus,
for $t<t_{\rm nr}$ we have $v_{\rm DM}\sim1$, $\rho_{\rm
DM}\propto a^{-4}$ and $\rho\simeq\rho_{\gamma}\propto a^{-4}$,
implying that $\lambda_{\rm J}^{({\rm c})}\propto a^2$ and $M_{\rm
J}^{({\rm c})}\propto a^2$. In the interval between $t_{\rm nr}$
and $t_{\rm dec}$ the key variables evolve as $v_{\rm DM}\propto
a^{-1/2}$ (recall that ${\rm T}_{\rm DM}\simeq{\rm T}_{\gamma}$
until $t_{\rm dec}$), $\rho_{\rm DM}\propto a^{-3}$ and
$\rho\simeq\rho_{\gamma}\propto a^{-4}$. As a result,
$\lambda_{\rm J}^{({\rm c})}\propto a^{3/2}$ and $M_{\rm J}^{({\rm
c})}\propto a^{3/2}$. After the particles have decoupled ${\rm
T}_{\rm DM}\neq{\rm T}_{\gamma}$, which means that $v_{\rm
DM}\propto a^{-1}$. At the same time $\rho_{\rm DM}\propto a^{-3}$
and $\rho\simeq\rho_{\gamma}\propto a^{-4}$, ensuring that
$\lambda_{\rm J}^{({\rm c})}\propto a$ and $M_{\rm J}^{({\rm
c})}={\rm constant}$. After equilibrium $v_{\rm DM}\propto a^{-1}$
and $\rho\propto\rho_{\rm DM}\propto a^{-3}$, implying that
$\lambda_{\rm J}^{({\rm c})}\propto a^{1/2}$ and $M_{\rm J}^{({\rm
c})}\propto a^{-3/2}$ In short, the Jeans mass of cold thermal
relics evolves as
\begin{equation}
M_{\rm J}^{({\rm c})}\propto\left\{\begin{array}{l}
a^2\hspace{20mm} z>z_{\rm nr}\,,\\
a^{3/2}\hspace{17mm}z_{\rm nr}>z>z_{\rm dec}\,,\\ {\rm
constant}\hspace{10mm}z_{\rm dec}>z>z_{\rm eq}\,,\\
a^{-3/2}\hspace{15mm}z_{\rm eq}>z\,.
\end{array}\right.  \label{cJm}
\end{equation}
Accordingly, the maximum value for ${\rm M}_{\rm J}^{({\rm c})}$
corresponds to species with $t_{\rm dec}=t_{\rm eq}$. In other
words, the sooner the particles cease being relativistic and
decouple the smaller the associated maximum Jeans mass. Typically
$(M_{\rm J}^{({\rm c})})_{\rm max}\ll10^{12}~{\rm M}_{\odot}$.

\subsection{Evolution of the Hubble mass}
By definition
\begin{equation}
M_{H}={\textstyle{4\over3}}\pi\rho_{\rm
DM}\left(\frac{\lambda_{H}}{2}\right)^3\,, \label{nbHm}
\end{equation}
where $\rho_{\rm DM}$ is the energy density of the collisionless
species. For $t<t_{\rm nr}$ we have $\rho_{\rm DM}\propto a^{-4}$,
$\lambda_{H}\propto t\propto a^2$. During the interval $t_{\rm
nr}<t<t_{\rm eq}$ we have $\rho_{\rm DM}\propto a^{-3}$ and
$\lambda_{H}\propto t\propto a^2$. Finally, when $t_{\rm eq}<t$,
$\rho_{\rm DM}\propto a^{-3}$ and $\lambda_{H}\propto t\propto
a^{3/2}$. Overall the Hubble mass of the dark matter component
evolves as
\begin{equation}
M_{H}\propto\left\{\begin{array}{l}
a^2\hspace{17mm} z>z_{\rm nr}\,,\\
a^3\hspace{17mm}z_{\rm nr}>z>z_{\rm eq}\,,\\
a^{3/2}\hspace{14mm}z_{\rm eq}>z\,.
\end{array}\right.  \label{nbHme}
\end{equation}
Following definitions (\ref{nbJl}), (\ref{nbJm}) and (\ref{nbHm}),
one can easily verify that the Jeans mass and the Hubble mass are
effectively identical as long as the relic species are
relativistic.

\subsection{Dissipative effects}
The ideal fluid approximation for collisionless species holds on
sufficiently large scales only. On small scales, the free geodesic
motion of the particles will wipe out any structure. This process
is known as ``Landau damping" or ``free streaming". A proper
calculation of the damping scale associated to free streaming
requires integrating the collisionless Boltzmann equation. Here,
we will only obtain an estimate of the effect in the case of hot
thermal relics. To begin with, consider the coordinate (comoving)
distance travelled by a free streaming particle
\begin{equation}
x_{\rm FS}=\int_0^t\frac{v_{\rm DM}}{a}{\rm d}t\,, \label{fss}
\end{equation}
where $\ell_{\rm FS}=ax_{\rm FS}$ is the corresponding physical
(i.e.~proper) distance. Clearly, perturbations in the dark matter
component on scales smaller than $\ell_{\rm FS}$ will be wiped out
by free streaming. Integrating the above we find, for $t<t_{\rm
nr}$ when $v_{\rm DM}\sim1$ and $a\propto t^{1/2}$
\begin{equation}
x_{\rm FS}=\frac{t_{\rm nr}^{1/2}}{a_{\rm
nr}}\int_0^t\,t^{-1/2}{\rm d}t=\frac{2t_{\rm nr}^{1/2}}{a_{\rm
nr}}t^{1/2}\,,  \label{fss1}
\end{equation}
on using the relation $a=a_{\rm nr}(t/t_{\rm nr})^{1/2}$. The
above suggests that $\ell_{\rm FS}=2t$ by the time the species
have ceased being relativistic. During the interval $t_{\rm
nr}<t<t_{\rm eq}$, when $v_{\rm DM}\propto a^{-1}$ (i.e.~$v_{\rm
DM}=a_{\rm nr}/a$) and $a\propto t^{1/2}$, the integration gives
\begin{equation}
x_{\rm FS}=\frac{2t_{\rm nr}}{a_{\rm nr}}+ a_{\rm nr}\int_{t_{\rm
nr}}^t\,a^{-2}{\rm d}t=\frac{2t_{\rm nr}}{a_{\rm
nr}}\left[1+\ln\left(\frac{a}{a_{\rm nr}}\right)\right]\,,
\label{ffs2}
\end{equation}
where $2t_{\rm nr}/a_{\rm nr}$ is the coordinate free streaming
distance at $t_{\rm nr}$. The associated physical free streaming
scale is $\ell_{\rm FS}=ax_{\rm FS}=2t_{\rm nr}/a_{\rm
nr}[1+\ln(a/a_{\rm nr})]a$. Finally, after equipartition $v_{\rm
DM}\propto a^{-1}$ (i.e.~$v_{\rm DM}=a_{\rm nr}/a$ still) and
$a\propto t^{2/3}$. Thus, a further integration of Eq.~(\ref{fss})
gives
\begin{eqnarray}
x_{\rm FS}&=&\frac{2t_{\rm nr}}{a_{\rm
nr}}\left[1+\ln\left(\frac{a_{\rm eq}}{a_{\rm nr}}\right)\right]+
a_{\rm nr}\int_{t_{\rm nr}}^t\,a^{-2}{\rm d}t\nonumber\\
&=&\frac{2t_{\rm nr}}{a_{\rm nr}}\left[1+\ln\left(\frac{a_{\rm
eq}}{a_{\rm nr}}\right)\right]+ \frac{3a_{\rm nr}t_{\rm
eq}}{a_{\rm eq}^2}\left[1-\left(\frac{t_{\rm
eq}}{t}\right)^{1/3}\right]\nonumber\\ &=&\frac{2t_{\rm
nr}}{a_{\rm nr}}\left[1+\ln\left(\frac{a_{\rm eq}}{a_{\rm
nr}}\right)\right]+ \frac{3t_{\rm eq}}{a_{\rm
nr}}\left[1-\left(\frac{a_{\rm eq}}{a}\right)^{1/2}\right]\,,
\label{ffs3}
\end{eqnarray}
on using the relation $t_{\rm eq}a_{\rm nr}^2=t_{\rm nr}a_{\rm
eq}^2$. Thus, the total physical fee-streaming scale is
\begin{equation}
\ell_{\rm FS}=\left\{\frac{2t_{\rm nr}}{a_{\rm
nr}}\left[1+\ln\left(\frac{a_{\rm eq}}{a_{\rm nr}}\right)\right]+
\frac{3t_{\rm eq}}{a_{\rm nr}}\left[1-\left(\frac{a_{\rm
eq}}{a}\right)^{1/2}\right]\right\}a\,.  \label{pfss}
\end{equation}
At late times, when $a\gg a_{\rm eq}$, the above approaches its
maximum value
\begin{equation}
\ell_{\rm FS}\rightarrow \left(\ell_{\rm FS}\right)_{\rm max}=
\frac{t_{\rm nr}}{a_{\rm nr}}\left[5+2\ln\left(\frac{a_{\rm
eq}}{a_{\rm nr}}\right)\right]\,,  \label{ltpfss}
\end{equation}
where the scale factor has been normalized so that $a=1$ at
present. To obtain numerical estimates we need to identify the
epoch the species become non-relativistic. Assuming that the
transition takes place when $T\sim m_{\rm DM}/3$ we find (see
Padmanabhan (1993)~\cite{KT})
\begin{equation}
\left(\ell_{\rm FS}\right)_{\rm max}\simeq0.5\left(\frac{m_{\rm
DM}}{1~{\rm keV}}\right)^{-4/3}\left(\Omega_{\rm
DM}h^2\right)^{1/3}~{\rm Mpc}\,, \label{mfss}
\end{equation}
where $m_{\rm DM}$ is the mass of the collisionless particles in
units of 1~keV. Accordingly, the minimum scale that survives
collisionless dissipation depends crucially on the mass of the
dark matter species. For neutrinos, with $m_{\rm \nu}\simeq30$~eV
we find $(\ell_{\rm FS})_{\rm max}\simeq28$~Mpc and a
corresponding mass scale $(M_{\rm FS})_{\rm max}\sim10^{15}~{\rm
M}_{\odot}$. Note that a more accurate treatment, using the
Boltzmann equation, gives $\ell_{\rm FS}\simeq40$~Mpc. For a much
heavier candidate, say $m_{\rm DM}\simeq1$~keV, we find
$(\ell_{\rm FS})_{\rm max}\sim0.5$~Mpc and $(M_{\rm FS})_{\rm
max}\sim10^9~{\rm M}_{\odot}$. In general, the lighter the dark
matter species less power survives on small scales.

Cold thermal relics and non-thermal relics have very small
dispersion velocities. As a result, the maximum values of the
Jeans mass and of the free streaming mass are very law. In this
case, perturbations on all scales of cosmological interest grow
unimpeded by damping processes, although they suffer stagnation
due to the Meszaros effect until matter-radiation equality. After
recombination the potential wells of the collisionless species can
boost the growth of perturbations in the baryonic component on
scales of the order of $(M_{\rm J}^{({\rm i})})_{\rm
rec}\sim10^5~{\rm M}_{\odot}$.

\subsection{Scenarios, successes and shortcomings}
Historically, there have been two major non-baryonic structure
formation scenarios. The ``Hot Dark Matter'' (HDM) models, where
the dominant collisionless matter is in the form of hot thermal
relics, and the ``Cold Dark Matter'' (CDM) models in which the
baryonic component is either a cold thermal relic or a non-thermal
species. It should be emphasized that, at present, pure HDM models
are not considered viable and that the simplest CDM models are
struggling to survive. Here, we will outline the key features of
these two old adversaries and present some of the current
alternatives.

\subsubsection{Hot dark matter models}
Typical HDM scenarios involve thermal relics with $z_{\rm
nr}\simeq z_{\rm eq}$ and the best motivated candidate is a light
neutrino species with $m_{\nu}\sim10$~eV. The key feature of the
perturbation spectrum is the cutoff at $\ell_{\rm FS}\simeq40$~Mpc
due to the neutrino free streaming. Because of this, the first
structures to form have sizes of approximately $10^{15}\,{\rm
M}_{\odot}$ which corresponds to a supercluster of galaxies (see
Figure 2). Moreover, because the scale is very large collapse must
have occurred at relatively recent times (i.e.~at $z<3$). Thus, in
a universe dominated by hot thermal relics, structure formation
proceeds in a top-down fashion similar to the adiabatic baryonic
models. Perturbations on scales as large as $10^{15}\,{\rm
M}_{\odot}$ go no linear in a highly non-spherical way. As a
result, they collapse to one dimensional objects resembling the
pancake-like formations of the baryonic models (Zeldovich
(1970)~\cite{Ze}). Once the pancake forms and goes non-linear in
one of its dimensions, the baryons within start colliding with
each other and dissipate their energy. Thereby, the baryonic
component fragments and condenses into smaller galaxy-sized
objects. The neutrinos, however, do not collide neither with each
other nor with the baryons. They are therefore unable to dissipate
their energy and collapse into more tightly bound objects. They
remain less condensed forming what one might call a ``neutrino
halo'' around the baryons.

\begin{figure}
\epsffile{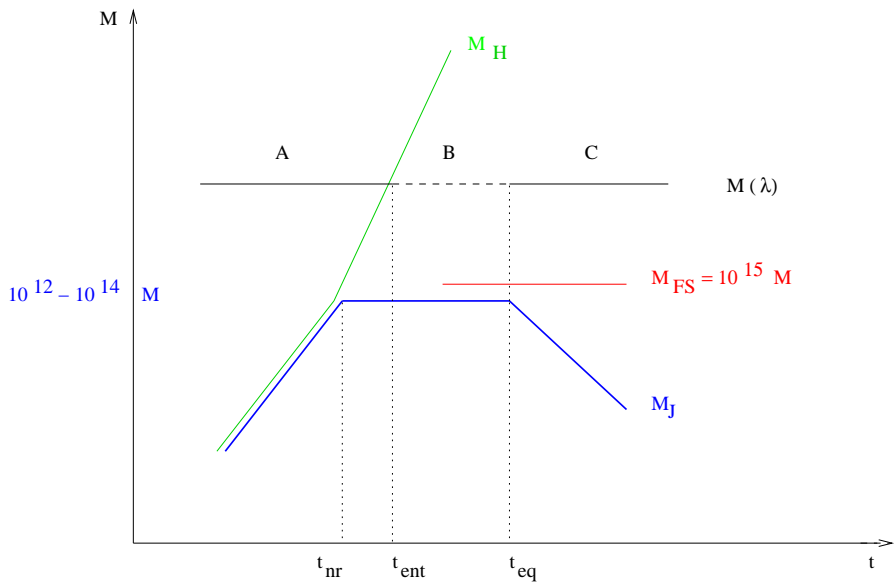}\caption{{\small Typical evolution of a
perturbed scale $M=M(\lambda)$ in HDM scenarios. During stage A
($t<t_{\rm ent}<t_{\rm eq}$) the mode grows as $\delta_{\rm
DM}\propto t$. Throughout stage B ($t_{\rm ent}<t<t_{\rm eq}$) the
perturbation is frozen in due to the Meszaros effect. Finally, in
stage C ($t_{\rm rec}<t$) the mode becomes unstable again and
grows as $\delta_{\rm b}\propto t^{2/3}$ Note that fluctuations
with size smaller than $M_{\rm FS}$ are wiped out by neutrino free
streaming.¥}} \label{fig: 2}
\end{figure}

Several groups have numerically simulated structure formation in a
neutrino dominated universe (e.g.~Centrella \& Mellott (1982);
White et al (1983)~\cite{CM}). In all simulations one notices a
cell-like structure, which reflects the damping scale due to
neutrino free streaming. The structure of these simulations on
scales larger than $10$~Mpc is qualitatively similar to the voids
and filaments seen in some of the redshift surveys. However, the
models have problems reproducing the small scale clustering
properties of galaxies. In particular, the HDM simulations can
agree with the observed galaxy-galaxy correlation function only if
the epoch of pancaking takes place at $z\simeq1$ or less. This
seems too late to account for the existence of galaxies with
redshift $z>1$ and of quasars with $z\simeq5$. A possible way out
is if the density field of the universe is traced by galaxy
clusters rather than by galaxies themselves. In this case the
mismatch between the predicted and the observed amount of
small-scale structure is alleviated. An additional problem of the
HDM scenarios is that the baryons are shock heated as they fall
into the pancakes. This might have raised the temperature to
levels that could have prevented the baryonic matter from
condensing or led to excessive x-ray production.

\subsubsection{Cold dark matter models}
Cold dark matter (CDM) candidates are cold thermal relics and
non-thermal relics with $z_{\rm nr}\ll z_{\rm eq}$. For such
particles the maximum damping scale is too small ($\ll1$~Mpc) to
be of any cosmological relevance. In the CDM models the important
feature is the weak growth experienced by perturbations between
horizon crossing and equipartition. It means that the density
contrast increases as we move to smaller scales, or that the
perturbation spectrum has more small-scale power. Note that the
shape of the HDM spectrum in not cosmologically important, as the
minimum scales to survive free streaming are way too large. Thus
in standard CDM scenarios the first objects to break away from the
background expansion have sub-galactic sizes ($<10^6~{\rm
M}_{\odot}$). These structures virialize through violent
relaxation (Lynden-Bell (1967); Shu (1978)~\cite{LB}) into
gravitationally bound configurations that resemble galactic halos.
At the same time the baryons can dissipate their energy and
condense further into the cores of these objects. As larger and
larger scales go non-linear, bigger structures form through tidal
interactions and mergers. Hence, according to the CDM scenario
structure forms in a bottom-up fashion analogous to that of the
isothermal baryonic models. Note that the ability of baryons to
dissipate allows objects of astrophysical size to condense out as
individual and discrete entities. The predicted properties of the
different kinds of galaxies that form this way appear in good
agreement with observations (Blumenthal et al (1984)~\cite{P1}).

The simplest CDM scenario, now known as the ``Standard CDM''
(SCDM) model, has critical density and contains only cold
collisionless species (Peebles (1982); Blumenthal et al (1984);
Davis et al (1985, 1992)~\cite{P1}). Major problems with these
scenarios are their failure of the simulations to reproduce the
observed galaxy-galaxy correlation function and the overproduction
of clusters when the modes are normalized with the COBE results.
One can attempt to save the SCDM model by removing the excess
small-scale power from the perturbation spectrum, an approach
known as tilting of the spectrum, (Vittorio et al (1988); Bond
(1992); Liddle et al (1992)~\cite{V}). This strategy, however,
usually has disastrous effects on the acoustic peaks of the CMB.
At the moment, the only way out for the SCDM model is a baryon
density substantially higher than the one predicted by standard
nucleosynthesis (White et al (1995b, 1996)~\cite{Wh}).

\subsubsection{Alternative options}
In an effort to salvage critical density, researchers have
abandoned the CDM hypothesis in favor of a mixture of cold and hot
dark matter particles (Bonommetto \& Valdarnini (1984); Fang et al
(1984); Shafi \& Stecker (1984); Holtzman (1989); Schaefer et al
(1989)~\cite{Bo}). Hybrid ``Cold-Hot Dark Matter'' (CHDM)
scenarios take advantage of the free streaming properties of their
hot collisionless component to reduce the small-scale power in the
perturbation spectrum. The original version had $\Omega_{\rm
CDM}=0.7$ and $\Omega_{\rm HDM}=0.3$ (Davis et al (1992); Schaefer
\& Shafi (1992, 1994); Taylor \& Rowan-Robinson (1992); Klypin et
al (1994)~\cite{D}), but the free streaming effect was too strong.
Currently, the preferred values for the hot dark matter
contribution lie near $\Omega_{\rm HDM}=0.2$ (Klypin et at (1995);
Pogosyan \& Starobinski (1995); Liddle et al (1996)~\cite{Kl}).
Note that CHDM models require a rather uncomfortable fine-tuning
to produce two particle species with similar cosmological
densities but very different masses.

If one wants to retain the CDM hypothesis, namely to assume cold
collisionless species only, the simple strategy is to reduce the
matter density. This shifts matter-radiation equality to a later
epoch, lengthening the period in which the small-scale modes have
their growth suppressed. In view of the very different
normalization of low-density universes with the COBE data, models
with $\Omega\simeq0.3$ ought to be easily distinguishable from
those with critical density. This is not the case, however, since
most other observations also require analogous renormalizations.

The natural way of keeping the low-density models compatible with
standard inflation is to introduce a cosmological constant, thus
obtaining the $\Lambda$CDM scenarios (Peebles (1984); Turner et al
(1984); Efstathiou et at (1990); Ostriker and Steinhardt
(1995)~\cite{P2}). Note that, at present, a non-zero cosmological
constant is favoured by the type-Ia suprenovae measurements
(Perlmutter et al, (1998); Schmidt et al (1998)~\cite{Pe}). The
presence of a non-zero $\Lambda$ means that one can capitalize on
the freedom to vary the rest of the parameters. At the moment it
appears that $\Omega\simeq0.3$ is the allowed lower limit on the
density, with the optimum value near $\Omega=0.3$. Note that one
can achieve extra freedom by assuming that the cosmological
constant is actually a decaying function of time, which is usually
referred to as ``quintessence'' (Coble et al (1997); Turner \&
White (1997); Caldwell et al (1998)~\cite{C})

An alternative way of shifting the time of equipartition is by
adding extra massless species in the model (Bardeen et al (1987);
Bond and Efstathiou (1991)~\cite{Ba2}). Recall that the present
density in relativistic particles is not known since the neutrino
background remains undetectable. Note, however, that the presence
of relativistic species is strongly constrained by
nucleosynthesis, which means that the extra relativistic energy
must be generated later on (Dodelson et al (1994); McNally \&
Peacock (1995); White et al (1995a)~\cite{Do}).

In all of the aforementioned models the perturbations are of the
adiabatic type. Scenarios based on pure isocurvature fluctuations
do not seem viable. However, an admixture of adiabatic and
isocurvature modes has not been excluded. In fact, the COBE data
do not seem to discriminate against an isocurvature component
(Stompor et al (1996)~\cite{St}).

Topological defects, such as cosmic strings, monopoles domain
walls and textures, provide a very different alternative to
structure formation (Vilenkin \& Shellard (1994); Hindmarsh \&
Kibble (1995)~\cite{K}). They have much more predictive power than
inflation, but they are also more technically demanding. Moreover,
current observations appear unfavorable to them (Allen et al
(1997); Pen et al (1997)~\cite{A}).

\section{Discussion}
The question of how the observed large-scale structure of the
universe developed and how galaxies were formed has been one of
the outstanding problems in modern cosmology. Looking back into
the past hundred years one sees three decisive moments in the
pursuit of the answer. The first milestone was the formulation of
general relativity which provided researchers with the theoretical
tool to probe the large-scale properties of the universe. Hubble's
observations manifesting the expansion of the universe was the
second decisive moment, as it forced cosmologists to break away
from the then prevailing concepts of a static and never-changing
cosmos. Finally, the discovery of the Cosmic Microwave Background
radiation by Penzias and Wilson established the Hot Big Bang
theory, an idea advocated several years earlier primarily by
Gamov. At last, cosmologists had a definite model within which
they could tackle the structure formation question. Pending future
observations, one could argue that the inflationary paradigm is
one additional milestone in our effort to understand the workings
of our universe. The recent supernovae measurements, which suggest
an accelerated universal expansion, could prove another very
decisive moment. Time will show whether they actually are.

Despite the problems and the uncertainties, cosmologists now
believe that all the structure that we observe around us today
originated from minute perturbations in a cosmic fluid that was
smooth to the accuracy of one part in ten thousand at the time of
recombination. Such tiny irregularities could have been triggered
by quantum fluctuations that were stretched out during the
inflationary expansion or by topological defects such as cosmic
strings for example. Given the current observational status,
inflation appears to be the strongest candidate. In these notes we
have set aside the question of the origin of the primeval
fluctuations. Our discussion focussed on the linear evolution of
these minute irregularities once the universe entered the
post-inflationary Hot Big Bang era, and on the physical processes
that could have affected them. When studying density
perturbations, one sooner or later encounters the reality that
baryon inhomogeneities cannot actually grow before recombination.
This fact, together with the extreme smoothness of the CMB
temperature, implies that baryonic fluctuations simply do not have
enough time to produce the observed structure. When one adds to
that the nucleosythesis constraints and the strong theoretical and
observational bias for spatial flatness, the once popular
baryon-dominated picture of the universe seems unsustainable. So,
if the baryons are not the dominant form of matter in our
universe, then what is? The answer to that might lie in high
energy physics theories. Theoretical physics provides a whole zoo
of supersymmetric, dark matter species that could bring $\Omega$
close to unity and also ``assist'' structure formation. The
attractive feature of collisionless matter is that perturbations
in its density start growing earlier than those in the baryonic
component. Thus, as soon as the baryons decouple from radiation,
they undergo a period of fast growth as they fall into the
potential wells of the collisionless species. This can improve the
final picture but unfortunately does not solve all the problems.
The key obstacle being that a single collisionless species does
not seem capable of fitting all the data. At best, one needs the
presence of one cold and one hot dark matter species in a
combination that will make the most of their advantages while
minimizing their shortcomings. The recent supernovae results,
suggesting that our universe might be dominated by some sort of
dark energy have added extra flavor to the whole situation. The
presence of a cosmological constant, or quintessence, means that
researchers have extra freedom when dealing with crucial
cosmological parameters, such as the age of the universe for
example. On the other hand, however, the idea that the majority of
the matter in our universe is in the form of some unknown exotic
species brings back some rather embarrassing memories from our
relatively recent past. A lot could be decided within the next few
years as we expect an influx of high quality data. The
``Boomerang'' and ``Maxima'' observations have added valuable
information which seems to favor the inflation based models. The
near future ``Probe'' and ``Planck'' satellite missions also
promise high precision data. For some researchers structure
formation is a story that is fast reaching its conclusion. The
future will show if they are right or just too hasty.

\section*{Acknowledgements}
I would like to thank Sotiris Bonanos, Marco Bruni, Theodosis
Christodoulakis, Peter Dunsby, George Ellis, Giorgos Kofinas, Roy
Maartens, Nikos Mavromatos Yiannis Miritzis and Manolis Plionis
for helpful discussions and comments. Special thanks to Spiros
Cotsakis and Lefteris Papantonopoulos for their invitation and
their hospitality during my stay in Karlovassi. Last, but not
least, I would like to thank the secretaries and the supporting
staff of the school for their help and kindness.\\

\noindent This work was partly supported by a Sida/NRF grant.


\begin{thebibliography}{99}
\bibitem[1]{J} J. Jeans: Phil Trans. {\bf 199A}, 49 (1902); J.
Jeans: {\em Astronomy and Cosmology}, Cambridge University Press
(1928).
\bibitem[2]{G} A.H. Guth: Phys. Rev. D {\bf 23}, 347 (1981); A.D.
Linde: Pys. lett. B {\bf 108}, 389 (1982); A. Albrecht and P.J.
Steinhardt: Phys. Rev. Lett. {\bf 48}, 1220 (1982).
\bibitem[3]{K} T.W.B. Kibble: J. Phys. {\bf A9}, 1387 (1976); A.
Vilenkin and E.P.S. Shellard: {\em Cosmic Strings and Topological
Defects}, Cambridge University Press (1994); M.B. Hindmarsh and
T.W.B. Kibble: Rep. Prog. Phys. {\bf 58}, 477 (1995).
\bibitem[4]{H} R. Harrison: Phys. Rev. D {\bf 1}, 2726 (1970); Y.B.
Zeldovich: Astron. Atrophys. {\bf 5}, 84 (1970).
\bibitem[5]{B} W.B. Bonnor: Mon. Not. R. Astron. Soc. {\bf 117},
104 (1957).
\bibitem[6]{L} E.M. Lifshitz: J. Phys. (Moscow) {\bf 10}, 116
(1946); E.M. Lifshitz and I.M. Khalatnikov: Adv. Phys. {\bf 12},
185 (1963).
\bibitem[7]{W} S. Weinberg: {\em Gravitation and
Cosmology}, Wiley (1972); L.D. Landau and E.M. Lifshitz: {\em The
Classical Theory of Fields}, Pergamon Press (1975); P.J.E.
Peebles: {\em The large Scale Structure of the Universe},
Princeton University Press (1980); W.H. Press and E.T. Vishniac:
{\em Astrophys. J.} {\bf 239}, 1 (1980).
\bibitem[8]{KT} E.W. Kolb and M.S. Turner: {\em The Early
Universe}, Addison-Wesley (1990); T. Padmanabhan: {\em Structure
Formation in the Universe}, Cambridge University Press (1993);
P.J.E. Pebles {\em Principles of Physical Cosmology}, Princeton
University Press (1993); P. Coles and F. Lucchin: {\em Cosmology:
The Origin and Evolution of Cosmic Structure}, Wiley (1995); J.A.
Peacock: {\em Cosmological Physics}, Cambridge University Press
(1999); A.R. Liddle and D.H. Lyth: {\em Cosmological Inflation and
Large-Scale Structure}, Cambridge University Press (2000).
\bibitem[9]{M} P. Meszaros: Astron. Astrophys. {\bf 37}, 225
(1974).
\bibitem[10]{S} R. Sachs: in {\em Relativity, Groups and Topology},
eds. De Witt and DeWitt, Gordon and Breach (1964); J. Stewart:
Class. Quantum Grav. {\bf 7}, 1169 (1990); M. Bruni, P.K.S. Dunsby
and G.F.R. Ellis: Astrophys. J. {\bf 395}, 34 (1992).
\bibitem[11]{Ba} J.M Bardeen: Phys. Rev. D {\bf 22}, 1882 (1980);
J.M. Bardeen, P.J. Steinhardt and M.S. Turner: Phys. Rev. D {\bf
28}, 679 (1983); U.H. Gerlach and U.K. Sengupta: Phys. Rev. D {\bf
18}, 1789 (1978); H. Kodama and M. Sasaki: Prog. Theo. Phys.
suppl. {\bf 78}, 1 (1984).
\bibitem[12]{EB} G.F.R Ellis and M. Bruni: Phys. Rev. D {\bf 40},
1804 (1989); S.W. Hwawking: Ap. J. {\bf 145}, 544 (1966); J.
Stewart and M. Walker: Proc. R. Soc. London {\bf A341}, 49 (1974);
D.W. Olson: Phys. Rev. D {\bf 14}, 327 (1976).
\bibitem[13]{EvE} G.F.R. Ellis and H. van Elst: in {\em Theoretical
and Observational cosmology}, edited by M. Lachi\`{e}ze-Rey
(Kluwer, Dordrecht, 1999)
\bibitem[14]{Z} Y.B. Zeldovich: Soviet Phys. Usp. {\bf 9}, 602
(1967)
\bibitem[15]{Si} J. Silk: Nature {\bf 215}, 1155 (1967); J. Silk:
Ap. J. {\bf 151}, 459 (1968); G. Efstathiou and J. Silk,: Fond.
Cosmic Phys. {\bf 9}, 1 (1983).
\bibitem[16]{Ze} Y.B. Zeldovich: Astrofisika {\bf 6}, 319 (1970).
\bibitem[17]{LB} D. Lynden-Bell: Mon. Not. R. Astron. Soc. {\bf
136}, 101 (1967); F.H. Shu: Astrophys. J. {\bf 225}, 83 (1978).
\bibitem[18]{CM} J. Centrella and A. Mellott: Nature {\bf 305}, 196
(1982); S.D.M. White, C. Frenk and M. Davis: Ap. J. {\bf 274}, L1
(1983); ibid {\bf 287}, 1 (1983).
\bibitem[19]{P1} P.J.E. Peebles: Astrophys. J. {\bf 263}, L1 (1982);
G.R. Blumenthal, S.M. Faber, J.R. Primack and M.J. Rees: Nature
{\bf 311}, 517 (1984); M. Davis, G. Efstathiou, C. Frenk and
S.D.M. White: Astrophys. J. {\bf 292}, 271 (1985); M. Davis, G.
Efstathiou, C. Frenk and S.D.M. White: Nature {\bf 356}, 489
(1992).
\bibitem[20]{V} N. Vittorio, S. Matarrese and F. Lucchin:
Atrophys. J. {\bf 328}, 69 (1988); J.R. Bond: in {\em Highlights
in Astronomy}, Vol. 9, {\em Proceedings of the IAU Joint
Discussion}, ed. J. Berjeron, Kluwer, Dordrecht (1992); A.R.
Liddle, D.H. Lyth and W. Sutherland: Phys. Lett. B {\bf 279}, 244
(1992).
\bibitem[21]{Wh} M. White, D. Scott, J. Silk and M. Davis: Mon.
Not. R. Astron. Soc. {\bf 276}, L69 (1995b); M. White, P.T.P.
Viana, A.R. Liddle and D. Scott: Mon. Not. R. Astron. Soc. {\bf
283}, 107 (1996).
\bibitem[22]{Bo} S.A. Bonommetto and R. Valdarnini: Phys. Lett. A
{\bf 103} 369 (1984); L.Z. Fang, S.X. Li and S.P. Xiang: Astron.
Astrophys. {\bf 140}, 77 (1984); Q. Shafi and F.W. Stecker: Phys.
Rev. Lett. {\bf 53}, 1292 (1984); J. Holtzman: Astrophys. J. Supp.
{\bf 71}, 1 (1989); R.K. Schaefer, Q. Shafi and F.W. Stecker:
strophys. J. {\bf 347}, 575 (1989).
\bibitem[23]{D} M. Davis, F. Summers and D. Schlegel: Nature {\bf
359}, 393 (1992); R.K. Schaefer and Q. Shafi: Nature {\bf 359},
199 (1992); R.K. Schaefer and Q. Shafi: Phys. Rev. D {\bf 49},
4990 (1994); A.N. Taylor and M. Rowan-Robinson: Nature {\bf 359},
396 (1992); A. Klypin, J. Holtzman, J.R. Primack and E. Reg\"{o}s:
Astrophys. J. {\bf 416}, 1 (1994).
\bibitem[24]{Kl} A. Klypin, S. Borgani, J. Holtzman and J.R.
Primack: Astrophys. J. {\bf 444}, 1 (1995); D.Y. Pogosyan and A.A.
Starobinski: Astrophys. J. {\bf 447}, 465 (195); A.R. Liddle, D.H.
Lyth, R.K. Shaefer, Q. Shafi and P.T.P. Viana: Mon. Not. R.
Astron. Soc. {\bf 281}, 531 (1996).
\bibitem[25]{P2} P.J.E. Peebles: Astrophys. J. {\bf 284}, 439
(1984); M.S. Turner, G. Steigman and L.M. Krauss: Phys. Rev. Lett.
{\bf 52}, 2090 (1984); J.P. Ostriker and P.J. Steinhardt: Nature
{\bf 377}, 600 (1995).
\bibitem[26]{Pe} S. Perlmutter et al: Nature {\bf 391}, 51 (1998);
B.P. Schmidt et al: Astrophys. J. {\bf 507}, 46 (1998).
\bibitem[27]{C} K. Coble, S. Dodelson and J.A. Frieman: Phys. Rev.
D {\bf 55}, 1851 (1997); M.S. Turner and M. White: Phys. Rev. D
{\bf 56}, R4439 (1997); R.R. Caldwell, R. Dave and P.J.
Steinhardt: Phys. Rev. Lett. {\bf 80}, 1586 (1998).
\bibitem[28]{Ba2} J.M. Bardeen, J.R. Bond and G. Efstathiou:
Astrophys. J. {\bf 321}, 28 (1987); J.R. Bond and G. Efstathiou:
Phys. lett. B {\bf 265}, 245 (1991);
\bibitem[29]{Do} S. Dodelson, G. Gyuk and M.S. Turner: Phys. Rev.
Lett. {\bf 72}, 3754 (1994); S.J. Mcnally and J.A. Peacock: Mon.
Not. R. Astron. Soc. {\bf 277}, 143 (1995); M. White, G. Gelmini
and J. Silk: Phys. Rev. D {\bf 51}, 2669 (1995a).
\bibitem[30]{St} R. Stompor, K.M. G\'{o}rski and A.J. Banday:
Astrophys. J. {\bf 463}, 8 (1996).
\bibitem[31]{A} B. Allen, R.R. Caldwell, S. Dodelson, L. Knox,
E.P.S. Shellard and A. Stebbins: Phys. Rev. lett. {\bf 79}, 2624
(1997); U.L. Penn, U. Seljak and N. Turok: Phys. Rev. Lett. {\bf
79}, 1611 (1997).
\end{thebibliography}
\end{document}